\documentclass[conference, onecolumn]{IEEEtran}
\usepackage{epsfig,rotating,setspace,latexsym,amsmath,epsf,amssymb,bm}
\usepackage{cite,graphicx,color,hyperref}
\usepackage{booktabs}
\IEEEoverridecommandlockouts
\begin{document}

\sloppy
\title{Towards Optimal Secure Distributed Storage Systems with Exact Repair\vspace{-5pt}} 
 \author{ 
   \IEEEauthorblockN{Ravi Tandon$^{\dagger}$, SaiDhiraj~Amuru$^{*}$, T.~Charles Clancy$^{*, \dagger}$ and R.~Michael Buehrer$^{*}$}
\IEEEauthorblockA{$^{*}$Bradley Department of Electrical and Computer
Engineering\\
$^{\dagger}$Hume Center for National Security and Technology,\\
Virginia Tech, Blacksburg, VA USA\\
Email: \{tandonr, adhiraj, tcc, rbuehrer\}@vt.edu}\vspace{-15pt}}


\newtheorem{Theo}{Theorem}
\newtheorem{remark}{Remark}
\newtheorem{Lem}{Lemma}
\newtheorem{Cor}{Corollary}
\newtheorem{Def}{Definition}

\maketitle
\begin{abstract}
Distributed storage systems in the presence of a wiretapper are considered. A distributed storage system (DSS) is parameterized by three parameters $(n, k,d)$,  in which a file stored across $n$ distributed nodes, can be recovered from any $k$ out of $n$ nodes. This is called as the reconstruction property of a DSS.  If a node fails, any $d$ out of $(n-1)$ nodes help in the repair of the failed node so that the regeneration property of the DSS is preserved. For such a $(n,k,d)$-DSS, two types of wiretapping scenarios are investigated: (a) Type-I (node) adversary which  can wiretap the data stored on any $l<k$ nodes; and a more severe (b) Type-II (repair data) adversary which can wiretap the contents of the repair data that is used to repair a set of $l$ failed nodes over time. The focus of this work is on the practically relevant setting of \emph{exact} repair regeneration in which the repair process must replace a failed node by its exact replica.   We make new progress on several non-trivial instances of this problem which prior to this work have been open. 

The main contribution of this paper is the optimal characterization of the secure storage-vs-exact-repair-bandwidth tradeoff region of a $(n,k,d)$-DSS, with $n\leq 4$ and any $l<k$ in the presence of both Type-I and Type-II adversaries. While the problem remains open for a general $(n,k,d)$-DSS with $n>4$, we present extensions of these results to a $(n, n-1,n-1)$-DSS, in presence of a Type-II adversary that can observe the repair data of any $l=(n-2)$ nodes. 
The key technical contribution of this work is in developing novel information theoretic converse proofs for the Type-II adversarial scenario. From our results, we show that  in the presence of Type-II  attacks, the  \textit{only efficient} point in the storage-vs-exact-repair-bandwidth tradeoff is the MBR (minimum bandwidth regenerating) point. This is in sharp contrast to the case of a Type-I  attack in which the storage-vs-exact-repair-bandwidth tradeoff allows a spectrum of operating points beyond the MBR point. 
\end{abstract}

\section{Introduction}
A continuous rise in the volume of data managed across various systems, calls for new storage mechanisms that maintain this data reliably. 
Distributed storage is the default technique for storing data in all new generation applications. The data from a file is stored in a decentralized manner on several un-reliable nodes/disks that when collectively used are capable of recovering the entire file. To ensure robustness to disk failures, the simplest scheme is to replicate and store the data across several disks. For instance, the Google File System (GFS) and the Hadoop Distributed File System (HDFS) store $3$ copies of the data across several nodes \cite{Rashmi_Facebook}. 
While  replication is robust to failures, it is not a scalable strategy to store large volumes of data. As an alternative, erasure codes (for instance, Reed-Solomon codes) have been used by Facebook, OceanStore, RAID-6 \cite{Rashmi_Shah_Kumar} and others to introduce redundancy into the storage system. However, to repair a failed node, the entire file must be downloaded from the remaining alive nodes. Thus, the repair process for such codes can be excessively bandwidth intensive. 
The concept of regenerating codes for distributed storage was introduced in the seminal work by Dimakis \emph{et al.} \cite{Dimakis_Intro}. A typical distributed storage system (DSS) consists of $n$ storage nodes each with a storage capacity of $\alpha$ (symbols or units of data) such that the entire file of size $\mathcal{B}$ can be recovered by accessing any $k\leq n$ nodes. This is called as the reconstruction property of the DSS. Whenever a node fails, $d$ nodes (where $k\leq d\leq n$) participate in the repair process by sending $\beta$ units of data each. This procedure is termed as the regeneration of a failed node and $\beta$ is often referred to as the per-node repair bandwidth. 

In \cite{Dimakis_Intro}, the authors show that the in order to store a file of size $\mathcal{B}$, the parameters of a DSS must necessarily satisfy 
\begin{align}\label{eq:functional_repair_tradeoff}
\mathcal{B}\leq \sum_{i=0}^{k-1}\mathsf{min}\left(\alpha,(d-i)\beta\right). 
\end{align}
Thus, in order to store a file of size $\mathcal{B}$, there exists a fundamental tradeoff between $\alpha$ (storage) and $d\beta$ (total repair bandwidth). Furthermore, it was shown that the reconstruction-regeneration requirements of a DSS can be equivalently formulated as a multicasting problem over an appropriately defined graph. This revelation along with the celebrated result of network coding for multicasting over graphs \cite{Net_Cod} were used to show that this tradeoff is indeed achievable.  However, this  tradeoff is in general achievable only for the functional-repair case. In functional repair, a failed node is replaced by a new node such that the resulting DSS has the same reconstruction and regeneration capabilities as before. In particular, the content of the repaired node may not necessarily be identical to the failed node  even though the desirable properties of the DSS are preserved. 

In contrast to functional repair,  exact repair regeneration requires the repair process to replace a failed node with an identical new node. Exact repair is appealing for many practical applications where the data has to be stored intact. The file recovery process is also easier in this case compared to the functional repair scenario since the file reconstruction procedure need not change whenever a failed node is replaced.  While characterizing the storage-vs-bandwidth tradeoff for the case of exact repair remains a challenging open problem in general, two extreme points of this tradeoff (depending on whether $\alpha$ or $\beta$ is minimized first) have been studied extensively. They are the minimum storage regenerating (MSR) and the minimum bandwidth regenerating (MBR) points for which the explicit exact-repair regenerating codes have been developed (see \cite{Shah_Rashmi_NonAchievable}, \cite{Cadambe_Jafar} and references therein).  Beyond these results, Tian \cite{Chao_Tian} has recently characterized the optimal exact-repair tradeoff for the $(4,3,3)$-DSS 
where it has been shown that there is a gap between the optimal tradeoffs for functional repair and exact repair. 

\begin{figure}[t]
\centering
\begin{minipage}[b]{0.45\columnwidth}
\centering
\includegraphics[width=1.00\columnwidth]{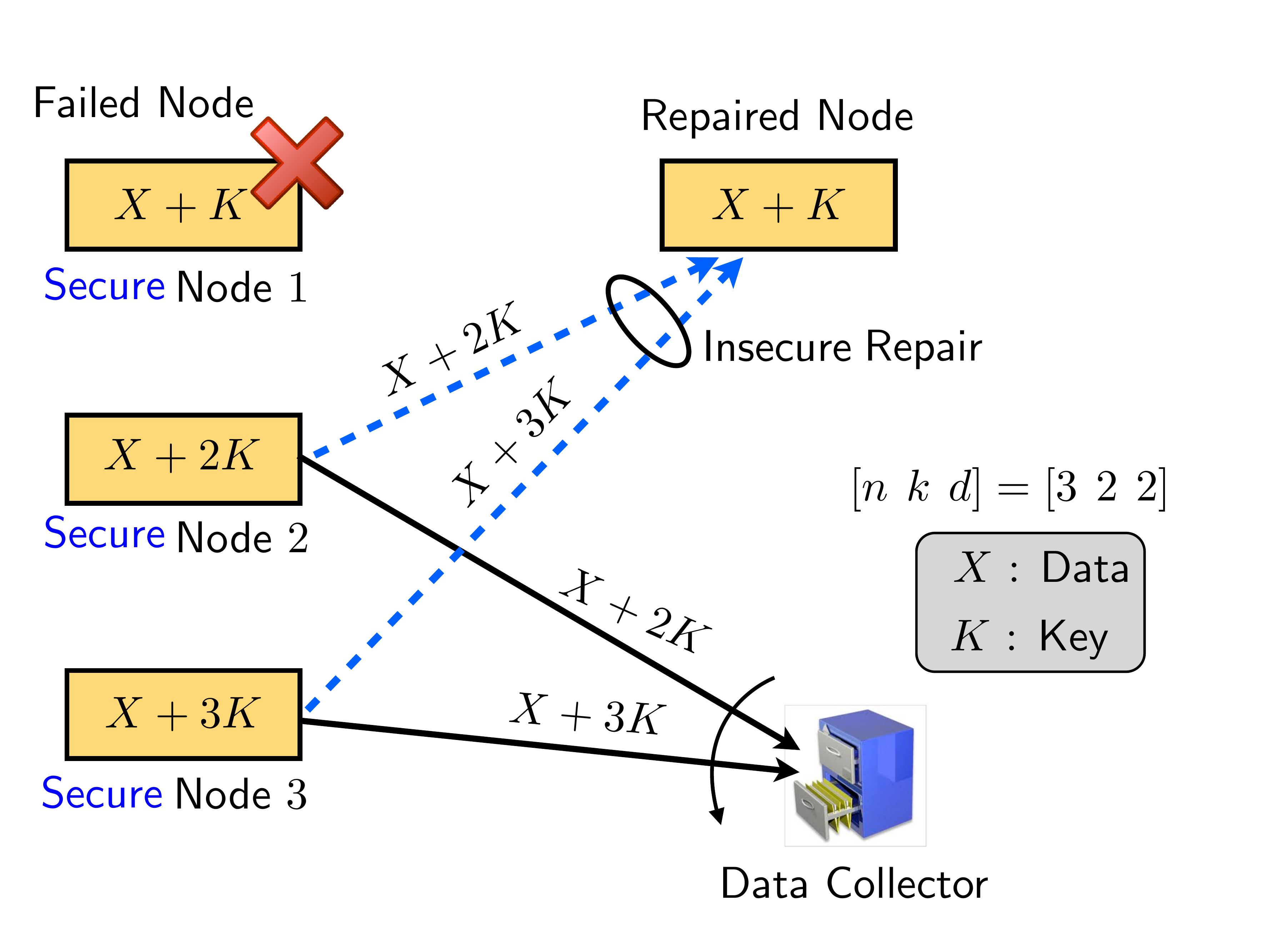}
\caption{$(3,2,2)$-DSS with Type-I (node) security.} 
\label{fig:Intro}
\end{minipage}
\begin{minipage}[b]{0.45\columnwidth}
\centering
\includegraphics[width=1.00\columnwidth]{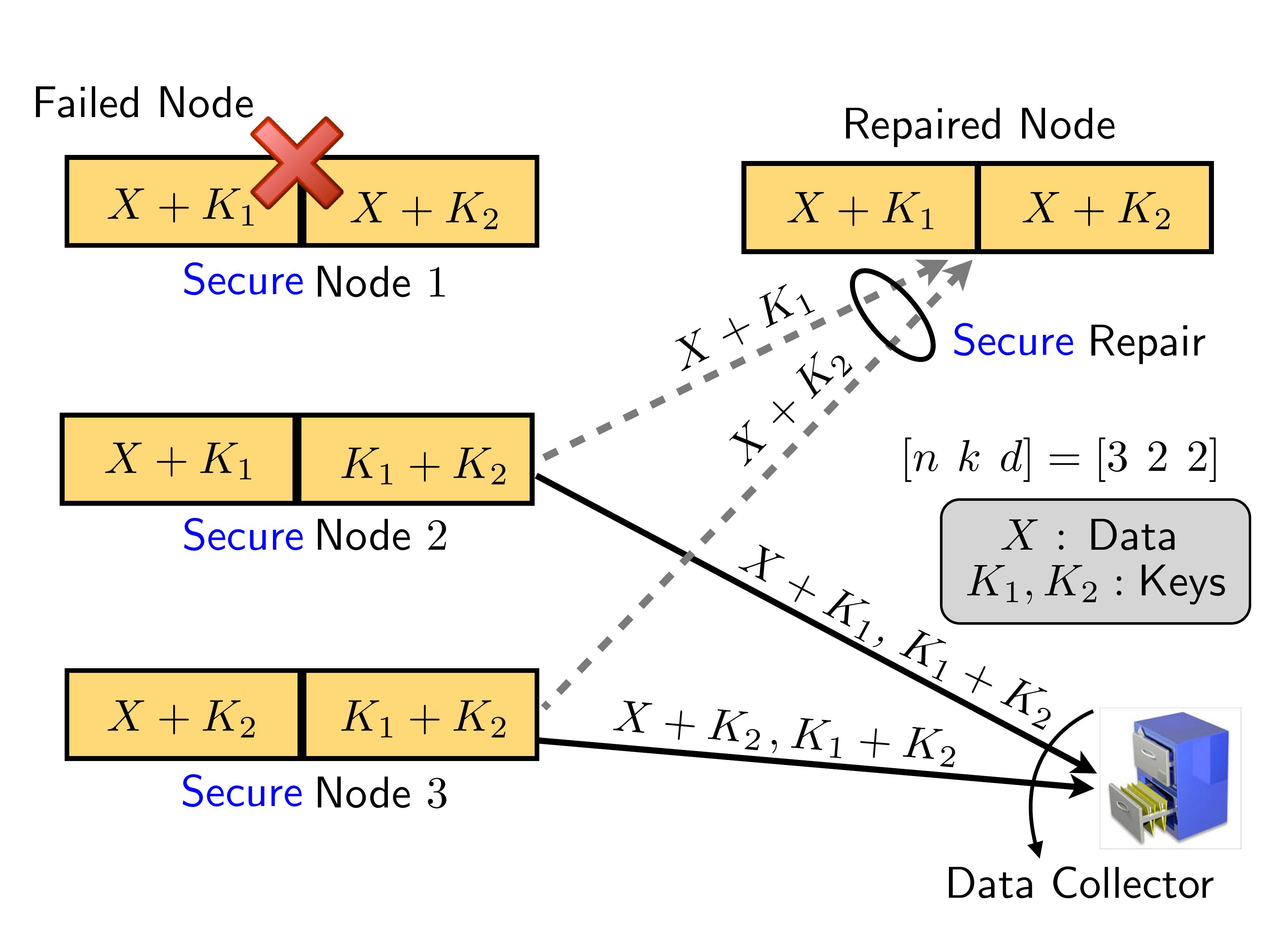}
\caption{$(3,2,2)$-DSS with Type-II (repair) security.} 
\label{fig:Intro2}
\end{minipage}
\vspace{-15pt}
\end{figure}
%

Sensitive data such as personal, health and financial records are increasingly being stored in a DSS. Securing such data from adversaries/eavesdroppers is necessary to ensure data secrecy for the users. Hence a DSS should be secure apart from satisfying the reconstruction and regenerating requirements. 
Two types of eavesdropping attacks can potentially occur in a DSS, (a) Type-I attack, in which the eavesdropper can wiretap the storage contents of $l$ nodes in the DSS and (b) Type-II attack, in which the eavesdropper wiretaps the contents of the repair data (and thereby the storage content as well) of $l$ nodes in the DSS. Throughout this paper, we assume that $l<k$ since $k$ is the minimum number of nodes required to reconstruct the entire file of size $\mathcal{B}$. Else, if $l\geq k$, the eavesdropper can recover the file by using the reconstruction property of the DSS.  
Encryption and decryption i.e., cryptographic approaches to ensure data security have been deemed to offer less secrecy compared to other information/coding theoretic approaches \cite{Rawat_Vishwanath}. Further, handling secure key management issues in a DSS are more complicated compared to developing information theoretically secure codes that offer the desired level of secrecy. 
 
The concept of security in a DSS against Type-II attacks was introduced in \cite{Pawar_Rouayheb} where, linear codes that achieve the optimal tradeoff region in the bandwidth limited regime were proposed. In the context of DSS, secure exact repair regenerating codes are beneficial as the functional repair process may reveal additional information such as coding coefficients that are used in the process of regenerating a functionally equivalent node \cite{Ozan_Rawat}. Optimal exact repair codes that are secure against eavesdropping have been explored for the MSR and MBR points in \cite{Shah_Rashmi_Secure}-\hspace{-0.5pt}\cite{Sreechakra_Rouyaheb}. 
The codes developed in \cite{Shah_Rashmi_Secure} achieve the MBR point for all $(n,k,d)$ configurations with any $l <k$. The MSR code in \cite{Shah_Rashmi_Secure} was shown to be optimal for Type-I attacks while it was shown to be optimal for Type-II attacks with $l=1$ in \cite{Rawat_Vishwanath}. The maximum file size $\mathcal{B}$ that can be securely stored using linear MSR codes with exact repair was studied in \cite{Sreechakra_Rouyaheb} and proved to be optimal for $d=n-1$. 

As an example, consider the  $(3,2,2)$-DSS and consider the setting in which the wiretapper can only read the contents of any $l=1$ node (Type-I adversary). For this setting, a secure exact repair code for this problem is shown in Fig. \ref{fig:Intro}. When the $1$st node fails, the other nodes send their data contents to enable its repair. Using these two data symbols, the $1$st node can recover its initial data contents (since two symbols can be recovered using two linearly independent combinations) thus satisfying the exact repair requirements. Since the eavesdropper is unaware of the secure key $K$, it cannot recover the data symbol $X$ by wiretapping on any single node in the DSS. 
%

Now, consider the same $(3,2,2)$-DSS in the presence of a more powerful Type-II adversary that can observe the repair data of any $l=1$ node. It is clear that the code in Figure \ref{fig:Intro} will leak out the entire data $X$ during the repair of node $1$, as the  wiretapper can use $X+2K, X+3K$, ($2$ linear combinations of $2$ symbols) to recover $X$.  Thus, a different secure exact repair DSS that can handle a Type-II adversary is shown in Fig.~\ref{fig:Intro2}. It is seen that the storage capacity should be increased to $\alpha=2$ in this case compared to the Type-I attack (where $\alpha=1$). Since the eavesdropper is not aware of the keys $K_1$ and $K_2$, it cannot get any information about the message symbol $X$ even if it observes the repair data of any one node. 

From these schemes, the secure repair scheme in Figure \ref{fig:Intro2} requires higher storage per-node ($\alpha=2$) compared to the secure node scheme in Figure \ref{fig:Intro} ($\alpha=1$ per-node). Thus, a natural question arises: does there exist a DSS  with a smaller storage per node $\alpha<2$ that can store a file of size $\mathcal{B}=1$, with repair bandwidth $\beta=1$ while still preserving the security of the repair process? Or, equivalently, is there any other more efficient tradeoff pair $(\alpha^{'},\beta^{'})$ than $(\alpha,\beta)=(2,1)$ that can store a file of size $\mathcal{B}=1$ while ensuring secure exact repair ?
 
In this paper, we answer this question in the negative through an information theoretic converse proof that shows that if secure and exact repair requirements are imposed, then the storage per-node cannot be smaller than $\alpha=2$ and hence the scheme in Figure \ref{fig:Intro2} is optimal. Equivalently, our result shows that the \textit{only efficient} point in the $(\alpha, \beta)$ tradeoff for this example is the MBR point which corresponds to $\alpha=d\beta$. This result  is extended and generalized for any $(n,n-1,n-1)$-DSS and Type-I/Type-II adversaries for $l=(k-1)=(n-2)$.

We next summarize the main contributions of this paper. 
\begin{enumerate}
\item We characterize and prove the optimal exact repair region that is achievable under Type-I and Type-II attacks for the $(n,k,d)$-DSS with $n\leq 4$ and $l<k$. Specifically, we look at $(3,2,2)$-DSS with $l=1$, $(4,2,3)$-DSS with $l=1$ and $(4,3,3)$-DSS with $l=1,2$. The main contributions of this paper are the novel converse proofs that characterize the strorage-vs-bandwidth tradeoff region against Type-II attacks for these distributed storage systems. 
\item For the (4,3,3)-DSS with $l=1$ and Type-I attack, 
we leverage upon the characterization of the exact repair region obtained in the absence of an eavesdropper in \cite{Chao_Tian}. Specifically, we introduce the additional security constraint in order to obtain the secure storage-vs-bandwidth tradeoff region in this case. Further, a novel code construction for this storage and bandwidth efficient region of the tradeoff curve is also introduced. This code construction also achieves the optimal exact repair region for $(4,3,3)$ in the absence of any eavesdroppers and is shown to be read efficient (in terms of the number of disk reads required to repair/recover a file \cite{PiggyBacking}) when compared to the code given in \cite{Chao_Tian}. 
\item These results are extended and proved for a general $(n,n-1,n-1)$-DSS when any $l=n-2$ nodes are compromised in the DSS. 
\end{enumerate}

The rest of this paper is organized as follows. In Section~\ref{System_Model} we describe the system and the eavesdropper's model. The main theorems that describe the optimal achievable storage-bandwidth tradeoff regions are given in Section~\ref{Theorems} and are proved in the Appendix. The intuition behind the converse proofs is explained in Section~\ref{Intuition}. The coding schemes that achieve these optimal regions are presented in Section~\ref{Achievability}. Finally conclusions are drawn in Section~\ref{Conclusion}\footnote{Parts of this paper will be presented at ITA-2014 \cite{ITA2014} and ICC-2014 \cite{ICC2014}.}.

\section{System Model}\label{System_Model}
A $\left(n,k,d,\alpha,\beta,\mathcal{B}\right)$-DSS  consists of $n$ storage nodes that store a file $F$ of size 
$\mathcal{B}$ across $n$ nodes, with each node capable of storing up to $\alpha$ units of data.  
A data collector can connect to any $k<n$ nodes and must be able to reconstruct the entire file $F$. This is known as the MDS property of the DSS \cite{Dimakis_Intro}. We focus on the scenario of single node failures in which at any given point any one node in the system could fail. For the repair of a failed node, any $d$ out of the remaining ($n-1$) \textit{alive} nodes can be accessed by downloading up to $\beta\leq \alpha$ symbols to repair the failed node. The parameter $d\beta$ is referred to as the total repair bandwidth.  

From an information theoretic perspective, the goal is to store a file $F$, whose entropy is $\mathcal{B}$, i.e.,
\begin{align}
H(F)=\mathcal{B}. 
\end{align}
We next introduce random variables corresponding to the data stored in the nodes and the random variables used in the file recovery and the repair process. Let $W_{i}$ denote the content that is stored at node $i$, for $i=1,2\ldots,n$. Hence,  the storage constraint implies
\begin{align}
H(W_{i})&\leq \alpha, \quad i=1,2,\ldots,n.\label{SM:storage}
\end{align}
Due to the MDS property (i.e., the file $B$ must be reconstructed from any $k\leq n$ nodes), we also have
\begin{align}
H\left(F|W_{\{k\}}\right)=0,\label{SM:regen}
\end{align}
where $W_{\{k\}}$ is the data stored in any subset of $k$ storage nodes. Next, we consider the repair of a failed node $j$ from any $d$ remaining nodes.  We denote the data sent by node $i$ to repair node $j$ by $S_{ij}$. Due to the repair bandwidth constraint, we have 
\begin{align}
&H(S_{ij})\leq\beta,\label{SM:repairbw}
\end{align}
and for \textit{exact repair} of node $j$ from the repair data of $d$ nodes, we also have
 \begin{align}
H(W_j|S_{r_{1}j}, S_{r_{2}j}, \ldots, S_{r_{d}j})=0,\ \{r_i\}_{i=1}^d\in [1,n]\neq j.\label{SM:repair}
\end{align}
Finally, we note that any repair data sent by node $i$, i.e., $S_{ij}$ is a function of the data stored in node $i$, i.e., $H(S_{ij}|W_{i})=0$.

\subsection{Adversarial Models}
\begin{figure}[t]
\centering
\begin{minipage}[b]{0.45\columnwidth}
\centering
\includegraphics[width=1.05\columnwidth]{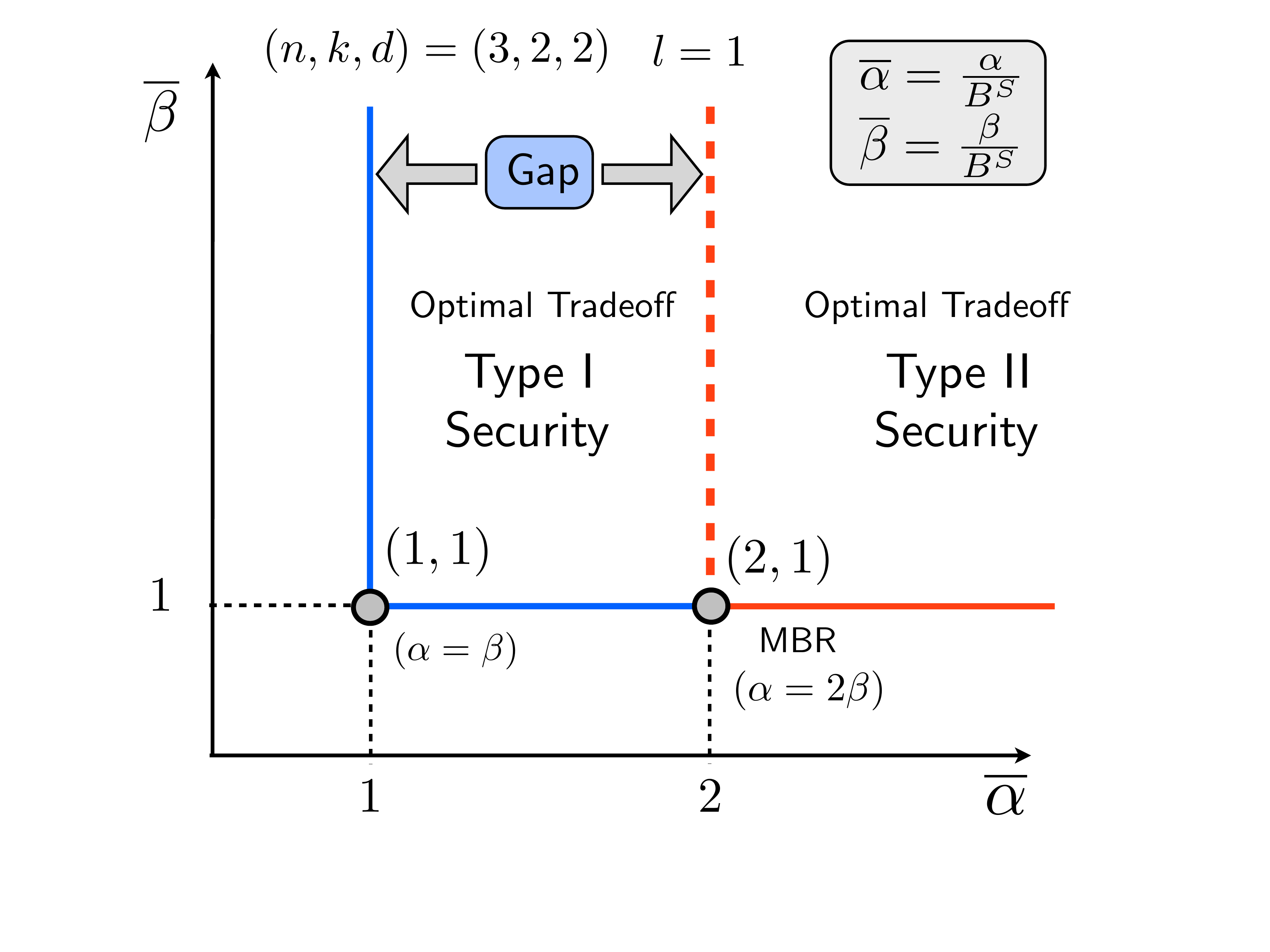}
\vspace{-32pt}
\caption{Secure $(\alpha,\beta)$ tradeoff for $(3,2,2)$-DSS and $l=1$.}
\label{fig:Theorem322}
\end{minipage}
\begin{minipage}[b]{0.45\columnwidth}
\centering
\includegraphics[width=1.05\columnwidth]{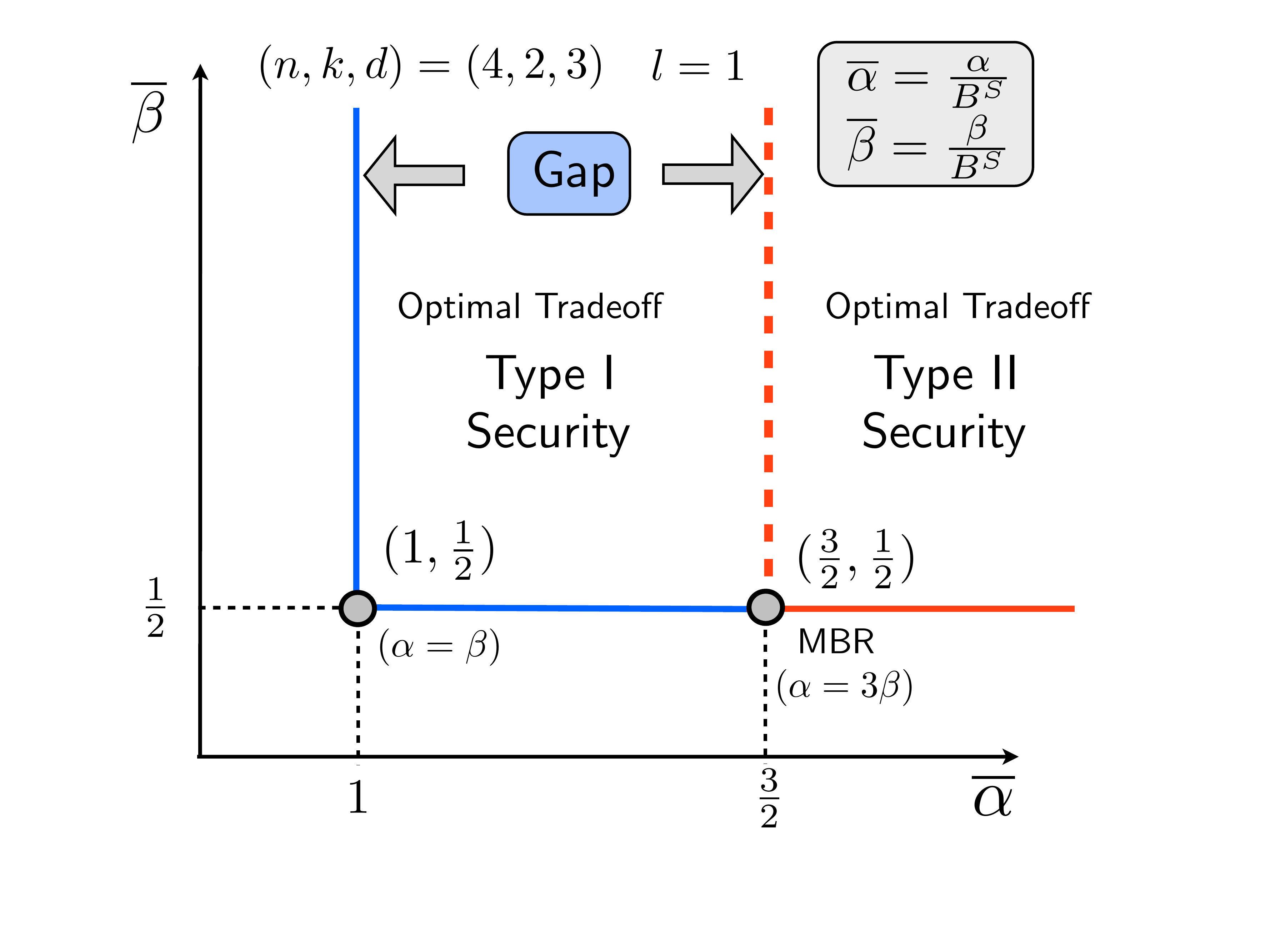}
\vspace{-32pt}
\caption{Secure $(\alpha,\beta)$ tradeoff for $(4,2,3)$-DSS and $l=1$. }
\label{fig:Theorem423L1}
\end{minipage}
\vspace{-5pt}
\end{figure}

We next formalize the Type-I and Type-II security constraints, each corresponding to different capabilities of the wiretapping adversary. 
\begin{itemize}
\item Type-I (node) security: in this setting, the wiretapper can read the storage content of any $l<k$ nodes.
Hence, we require that the information leakage by revealing the data of any $l$ storage nodes must be zero, i.e.,
\begin{align}
I\left(F; W_{\{l\}}\right)=0,\label{SM:TypeI}
\end{align}
where $W_{\{l\}}$ is the data stored in any subset of $l$ nodes.

\item Type-II (repair) security: in this setting, the wiretapper can read the repair data of any $l<k$ nodes over time.  Hence, for the repair of any  $l$ nodes to be secure, we require
\begin{align}
I\left(F; S_{n_{1}}, S_{n_{2}}, \ldots, S_{n_{l}}\right)=0,\label{SM:TypeII}
\end{align}
where $S_{n_{i}}$ is the data (downloaded from $d$ repair nodes) used in the repair of node $n_{i}$. Furthermore, it is worth noting that any DSS that is secure under Type-II secrecy constraint is also secure under Type-I secrecy constraint but the reverse statement is not true in general.  
\end{itemize} 
To illustrate these constraints, consider the $(n,k,d)= (3,2,2)$-DSS. Then, the constraints regarding file-regeneration and exact repair are as follows:

\noindent File Regeneration:
\begin{align}
H(F|W_{1}, W_{2})=H(F|W_{1}, W_{3})= H(F|W_{2}, W_{3})=0.\nonumber
\end{align}
Exact Repair:
\begin{align}
\hspace{-4pt}H(W_{1}|S_{21}, S_{31})= H(W_{2}|S_{12}, S_{32})=H(W_{3}|S_{13}, S_{23})=0\nonumber.
\end{align}
Repair data functionality:
\begin{align}
\hspace{-4pt}H(S_{12}, S_{13}|W_{1})=H(S_{21}, S_{23}|W_{2})=H(S_{31}, S_{32}|W_{1})=0\nonumber.
\end{align}
For this example, the parameter $l$ can be either $0$ and $1$. The non-trivial case is $l=1$, for which the Type-I constraints can be written as:
\begin{align}
I(F;W_{1})=I(F;W_{2})=I(F;W_{2})=0.\nonumber
\end{align}
whereas the Type-II security constraints are:
\begin{align}
\hspace{-15pt}I(F;S_{21}, S_{31})=I(F;S_{12}, S_{32})=I(F;S_{13}, S_{23})=0.\nonumber
\end{align}
We next define the Type I (respectively Type-II) secrecy capacity of a DSS as the maximum file size that can be stored
under the constraints placed on storage (\ref{SM:storage}), file regeneration (\ref{SM:regen}), repair bandwidth constraint (\ref{SM:repairbw}), exact repair (\ref{SM:repair}) and Type-I (respectively Type-II) secrecy constraint. Formally, we define the Type-I secrecy capacity as
\begin{align}
\mathcal{B}^{S}_{I}&=\max_{(\ref{SM:storage})- (\ref{SM:repair}), (\ref{SM:TypeI})}  H(F),
\end{align}
and the Type-II secrecy capacity as
\begin{align}
\mathcal{B}^{S}_{II}&=\max_{(\ref{SM:storage})- (\ref{SM:repair}), (\ref{SM:TypeII})}  H(F).
\end{align}
The study of distributed storage systems in the presence of a passive wiretapper was initiated in \cite{Pawar_Rouayheb}.  
It was shown that for any $(n,k,d)$-DSS with either Type-I or Type-II secrecy constraint (characterized by the parameter $l$), the following is an upper bound on the maximum secure file size $\mathcal{B}^{S}$: 
\begin{align}\label{eq:functional_repair_tradeoff_secure}
\mathcal{B}^S \leq \sum_{i=l}^{k-1}\mathsf{min}(\alpha,(d-i)\beta). 
\end{align} 
Intuitively, this result can be interpreted as follows. In the presence of a wiretapper, the maximum file size a DSS can store must necessarily reduce (compared to  \eqref{eq:functional_repair_tradeoff})  because $l$ nodes are wiretapped. 
Since $l$ nodes are compromised, at most $(k-l)$ nodes can help a data collector in recovering the entire file while keeping it secure from $l$ nodes. Hence the summation is over $(k-l)$ nodes as opposed to $k$ nodes \cite{Shah_Rashmi_Secure}. 
To illustrate by an example, consider the $(3,2,2)$-DSS with $l=1$ for which the upper bound in (\ref{eq:functional_repair_tradeoff_secure}) simplifies to  
\begin{align}
\mathcal{B}^{S}&\leq \min(\alpha, \beta).\label{explain322}
\end{align}
Normalizing (\ref{explain322}) throughout by $\mathcal{B}^{S}$, and defining  $\overline{\alpha}=\alpha/\mathcal{B}^{S}$, $\overline{\beta}=\beta/\mathcal{B}^{S}$, we can equivalently write
$1\leq \min(\overline{\alpha}, \overline{\beta})$.  This bound gives rise to a region in the $(\overline{\alpha}, \overline{\beta})$ plane and is shown in Fig.~\ref{fig:Theorem322} with the corner point $(\overline{\alpha}, \overline{\beta})=(1,1)$. Furthermore, it is not difficult to show that this tradeoff is achievable under Type-I security constraints by showing that it is possible to store a file of size $\mathcal{B}^{S}=1$ using $\alpha=1$ and $\beta=1$. The scheme is shown in Figure \ref{fig:Intro}. Hence, the upper bound in (\ref{explain322}) along with the scheme of Figure \ref{fig:Intro} imply that the optimal $(\alpha,\beta)$-tradeoff for $(3,2,2)$-DSS, $l=1$ in presence of a Type-I adversary is given by
\begin{align}
\mathcal{B}^{S}_{I}&\leq \min(\alpha, \beta).
\end{align}

As for the more stringent Type-II security is concerned, it has been shown by Shah \emph{et al} in \cite{Shah_Rashmi_Secure}, using the Product-Matrix framework at the MBR point ($\alpha=d\beta$), that it is possible to store a file of size $\mathcal{B}^{S}=1$ using 
$\alpha=2$ and $\beta=1$. This scheme (shown in Figure \ref{fig:Intro2}) shows that the following $(\alpha,\beta)$-tradeoff region is achievable:
\begin{align}
\mathcal{B}^{S}_{II}\leq \min\left(\frac{\alpha}{2}, \beta\right).
\end{align}
On the other hand, the bound in (\ref{explain322}) implies that the optimal $(\alpha,\beta)$-tradeoff region for the Type-II secrecy constraint is contained in:
\begin{align}
\mathcal{B}^{S}_{II}\leq \min(\alpha, \beta).
\end{align} 
Thus, there exists a significant gap between these two regions for the Type-II secrecy constraint. Furthermore, it is not clear whether the gap is due to the weakness of the upper bound or 
if there exists an improved achievable scheme that can close this gap. In this paper, we show that this gap is \textit{fundamental} to the secure exact repair problem and the upper bound can be improved. By deriving a novel information theoretic converse, 
we show that for the $(3,2,2)$-DSS, $l=1$ and Type-II secrecy constraint, we show that the optimal secure storage-repair-bandwidth tradeoff is given by
\begin{align}
\mathcal{B}^{S}_{II}&\leq \min\left(\frac{\alpha}{2}, \beta\right).
\end{align}

\section{Main Results}\label{Theorems}
In this section, we outline the main theorems that describe the optimal secure storage-vs-exact repair-bandwidth tradeoffs (in short referred to as $(\alpha, \beta)$-tradeoff region) under the exact repair requirement. 

\begin{Theo}\label{Theorem322}
The optimal $(\alpha,\beta)$-tradeoff regions for $(3,2,2)$-DSS with $l=1$ under exact repair are given by:
\begin{align}
\mathcal{B}^{S}_{I}&\leq \min(\alpha,\beta),\label{Type1322}\\
\mathcal{B}^{S}_{II}&\leq \min\left(\frac{\alpha}{2},\beta\right).\label{Type2322}
\end{align}
\end{Theo}

The converse proof for (\ref{Type1322}) follows directly from (\ref{eq:functional_repair_tradeoff_secure}) and is therefore omitted. 
The main contribution is the converse proof for the bound (\ref{Type2322}) which is given in Section \ref{Intuition}.

\begin{Theo}\label{Theorem4232}
The optimal $(\alpha,\beta)$-tradeoff regions for $(4,2,3)$-DSS with $l=1$ under exact repair are given by:
\begin{align}
\mathcal{B}^{S}_{I}&\leq \min(\alpha,2\beta)\label{Type14231}\\
\mathcal{B}^{S}_{II}&\leq \min\left(\frac{2\alpha}{3},2\beta\right).\label{Type24232}
\end{align}
\end{Theo}
The bound in \eqref{Type14231} follows directly from \eqref{eq:functional_repair_tradeoff_secure}. The converse proof for \eqref{Type24232} is given in Section~\ref{ConvProofTheorem4232} in the Appendix. 

\begin{Theo}\label{Theorem4331}
The optimal $(\alpha,\beta)$-tradeoff regions for $(4,3,3)$-DSS with $l=1$ under exact repair are given by:
\begin{align}
\mathcal{B}^{S}_{I}&\leq \min\left(\min(\alpha,2\beta)+ \min(\alpha, \beta), \frac{\alpha+6\beta}{3}\right)\label{OldType1}\\
\mathcal{B}^{S}_{II}&\leq \min\left(\alpha,3\beta\right).\label{Type24331}
\end{align}
\end{Theo}

The first term in \eqref{OldType1} is directly from the upper bound given in \eqref{eq:functional_repair_tradeoff_secure}. The converse proof for the second term in \eqref{OldType1} is one novel contribution of this paper. This proof is leveraged off the proof of exact repair given for the $(4,3,3)$-DSS without any secrecy constraints in \cite{Chao_Tian}. This is presented in Section~\ref{ConvProof43311} in the Appendix. A novel converse proof for the bound in \eqref{Type24331} under the Type-II constraints is given in Section~\ref{ConvProof43312} in the Appendix.

\begin{Theo}\label{Theorem4332}
The optimal $(\alpha,\beta)$-tradeoff regions for $(4,3,3)$-DSS with $l=2$ under exact repair are given by:
\begin{align}
\mathcal{B}^{S}_{I}&\leq \min(\alpha,\beta)\label{Type14331}\\
\mathcal{B}^{S}_{II}&\leq \min\left(\frac{\alpha}{3},\beta\right).\label{Type24332}
\end{align}
\end{Theo}
The bound in \eqref{Type14331} follows directly from \eqref{eq:functional_repair_tradeoff_secure}. Hence the proof is skipped. The converse proof for \eqref{Type24332} is given in Section~\ref{ConvProofTheorem4332} in the Appendix. 

\begin{Theo}\label{Theoremnnn}
The optimal $(\alpha,\beta)$-tradeoff regions for $(n,n-1,n-1)$-DSS with $l=n-2$ under exact repair are given by:
\begin{align}
\mathcal{B}^{S}_{I}&\leq \min(\alpha,\beta)\label{Typennn1}\\
\mathcal{B}^{S}_{II}&\leq \min\left(\frac{\alpha}{n-1},\beta\right).\label{Typennn2}
\end{align}
\end{Theo}

This is an extension of the Theorems~\ref{Theorem322} and \ref{Theorem4332} to a general $(n,n-1,n-1)$-DSS. This is the worst case scenario with respect to the $(n,n-1,n-1)$-DSS since $l=k-1$ nodes are compromised and hence the file size that can be stored securely is the minimum among all possible $l<k$ scenarios. Similar to the Theorems~\ref{Theorem322}-\ref{Theorem4332} the proof of \eqref{Typennn1} is omitted as it is derived from \eqref{eq:functional_repair_tradeoff_secure}. The proof for \eqref{Typennn2} is a novel extension of the converse proofs given for \eqref{Type2322} \eqref{Type24332} and is provided in the Appendix in Section \ref{ConvProofTheoremn}.

\begin{figure}[t]
\centering
\begin{minipage}[b]{0.45\columnwidth}
\centering
\includegraphics[width=1.05\columnwidth]{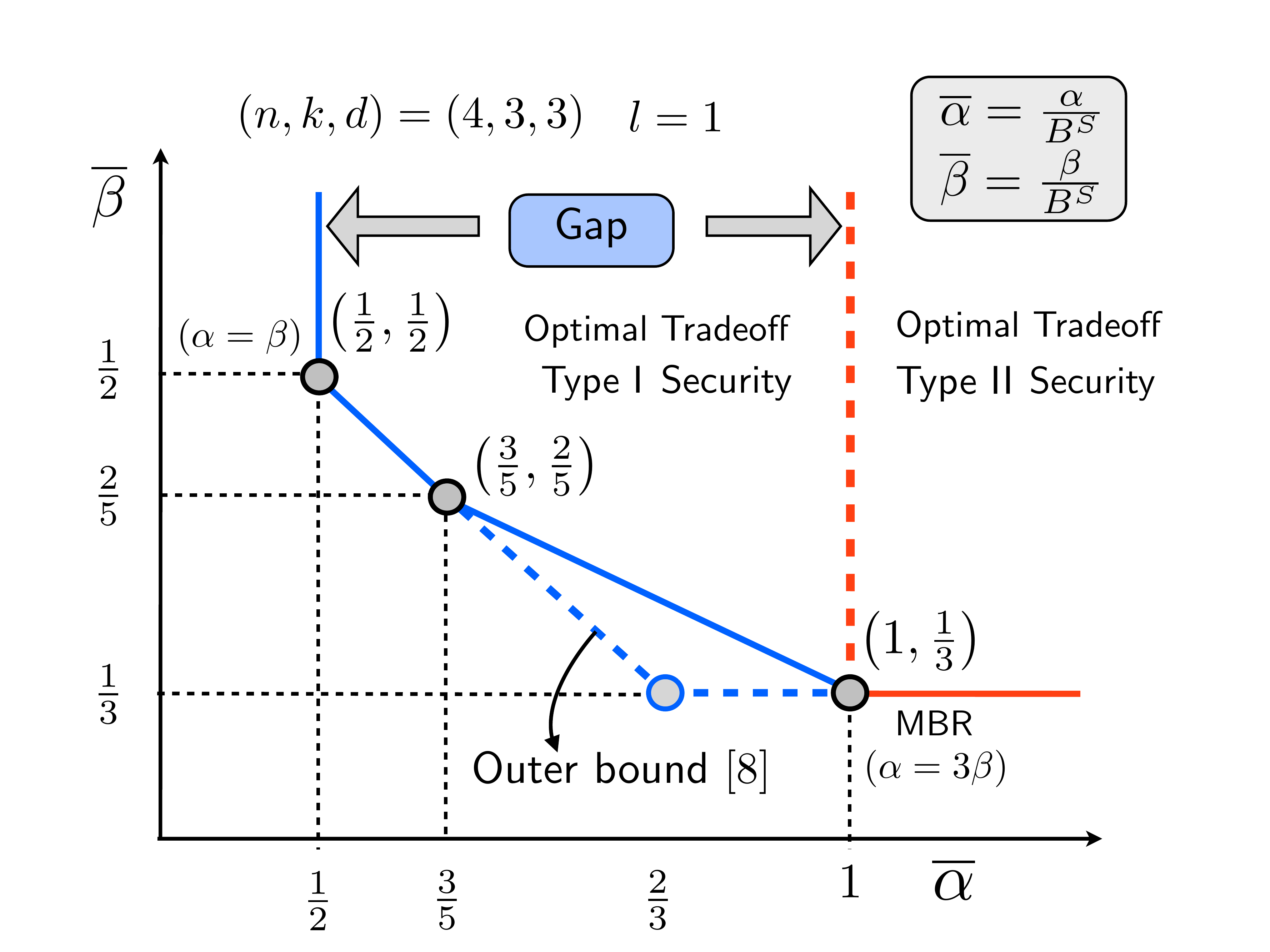}
\caption{Secure $(\alpha,\beta)$ tradeoff for $(4,3,3)$-DSS and $l=1$.} 
\label{fig:Theorem433}
\end{minipage}
\begin{minipage}[b]{0.45\columnwidth}
\centering
\includegraphics[width=1.05\columnwidth]{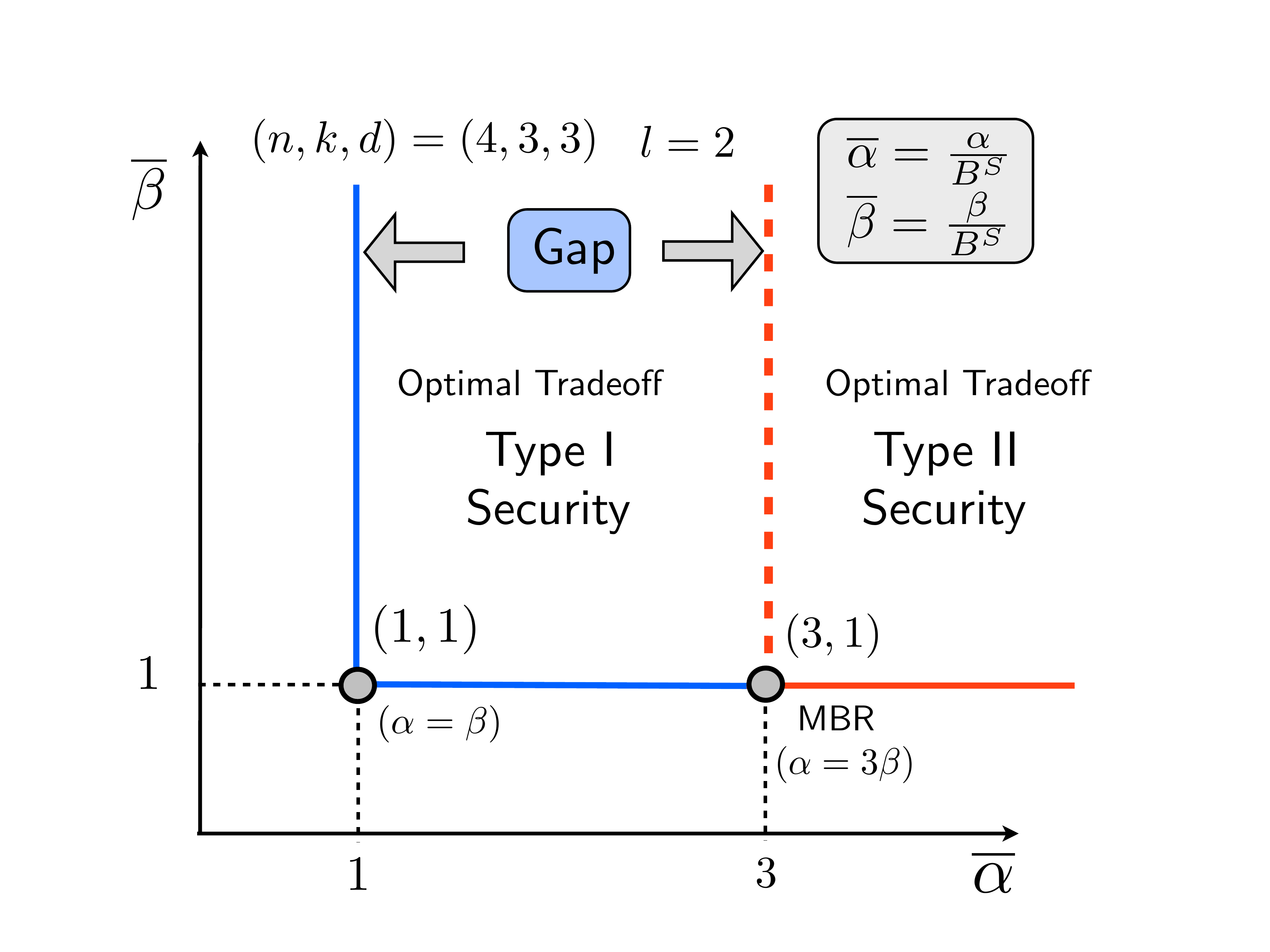}
\caption{Secure $(\alpha,\beta)$ tradeoff for $(4,3,3)$-DSS and $l=2$.} 
\label{fig:Theorem433L2}
\end{minipage}
\vspace{-5pt}
\end{figure}

%

\begin{figure}[t]
\centering
\includegraphics[width=0.495\columnwidth]{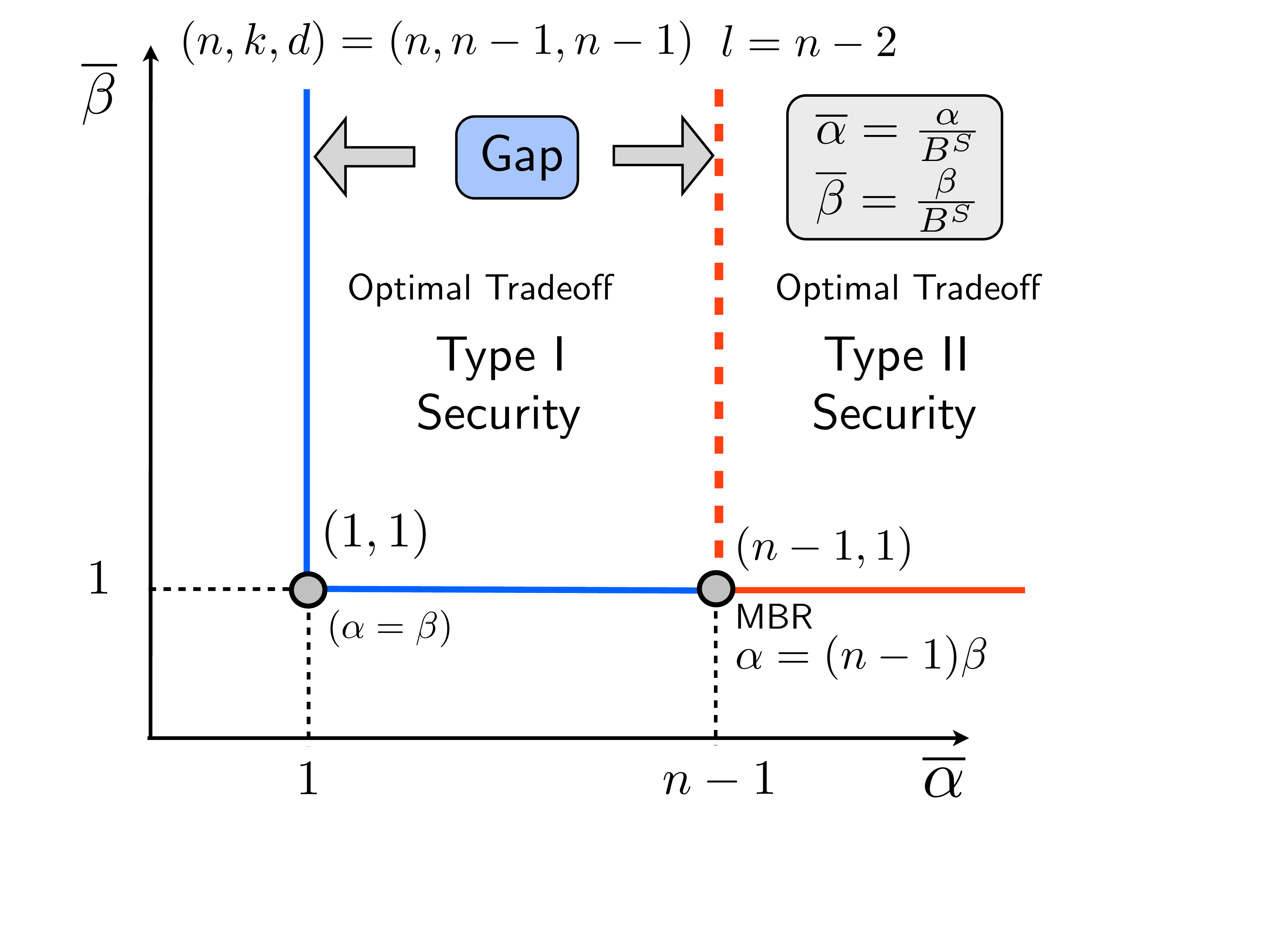}
\vspace{-28pt}
\caption{Secure $(\alpha,\beta)$ tradeoff for $(n,n-1,n-1)$-DSS and $l=n-2$. }
\vspace{-9pt}
\label{fig:TheoremnnnLn2}
\end{figure}

Figs.~\ref{fig:Theorem322}-\ref{fig:TheoremnnnLn2} show the optimal $(\alpha,\beta)$ tradeoff regions described by Theorems~\ref{Theorem322}-\ref{Theoremnnn}. From these theorems, it is seen that the only efficient point in the  optimal $(\alpha,\beta)$-tradeoff region for the Type-II constraints is the MBR point where $\alpha=d\beta$. However, this is different from the optimal tradeoff-region achievable under Type-I constraints as seen in these results. Thus there is a gap between the optimal regions achievable under these two constraints i.e., the file size that can be securely stored under Type-II constraints is lower than the file size achieved under Type-I constraint. 

Further notice that the gap between these two constraints with respect to the secure file size increases as $n$ increases in the DSS. For the Type-I attacks, when $l=n-2$ and $k=n-1$, only one node can securely store the file and hence the maximum file size that can be securely stored is given by \eqref{Typennn1} which is independent of $n$, while it decreases due to the wiretapping process as seen in \eqref{Typennn2} (a more stringent constraint on the DSS). 

\section{Proof of Theorem~\ref{Theorem322} and Intuition behind Converse Proofs}\label{Intuition}
In this section, we present the proof of the Theorem~\ref{Theorem322} i.e., $(3,2,2)$-DSS with $l=1$ in order to explain the intuition behind the novel converse proofs that establish the secure storage-vs-bandwidth repair tradeoff region against Type-II attacks. The converse proofs for the other Theorems are presented in the Appendix. 

Let $F$ be the random variable that denotes the file stored on the DSS. For the $(3,2,2)$-DSS with $l=1$, the bound in \eqref{eq:functional_repair_tradeoff_secure} (corresponding to Type-I attack) is given by 
\begin{align}
H(F)\leq \mathsf{min}(\alpha,\beta).
\end{align}
In the Type-I attack, the stored content of any $l=1$ node must not reveal any information about the file $F$. For instance, the content stored in node $1$ must not reveal any information about $F$, i.e., the DSS must satisfy $I(F;W_{1})=0$. Using such Type-I security constraints, one can readily show that $H(F)\leq \min(\alpha, \beta)$ as follows:
\begin{align}
H(F)&=H(F|W_1) \\
&=H(F,W_1)-H(W_1)\\
&\leq H(F,W_1,W_2)-H(W_1)\\
&=H(W_1,W_2) + \underbrace{H(F|W_{1}, W_{2})}_{=0 \text{ (file regeneration)}}-H(W_1) \label{2_nodes_give_file}\\
&= H(W_1, W_2)-H(W_1)\\
&= H(W_{2}|W_{1})
\end{align}
Next, the conditional entropy $H(W_{2}|W_{1})$ can be expanded in two distinct ways as follows:
\begin{enumerate}
\item[(a)] By bounding this term as
\begin{align}
H(W_{2}|W_{1})\leq H(W_{2})\leq \alpha,
\end{align}
gives the bound $H(F)\leq \alpha$.
\item[(b)] On the other hand, we can also bound this term as
\begin{align}
H(W_{2}|W_{1})&\leq H(W_{2},S_{32}|W_{1})\\
&= H(S_{32}|W_{1}) + \underbrace{H(W_{2}|S_{32}, W_{1})}_{=0 \text{ (repair of node 2)}}\\
&= H(S_{32}|W_{1})\\
&\leq H(S_{32})\leq \beta,
\end{align}
which gives the bound $H(F)\leq \beta$.
\end{enumerate}
While this bound $H(F)\leq \min(\alpha, \beta)$ is tight for the case of Type-I attack, it turns out to be strictly sub-optimal for the case of more severe Type-II attack. In particular, for the Type-II attack, 
in order to secure the DSS from an eavesdropper that wiretaps the repair process of any single node $(l=1)$, the DSS must satisfy the following three security constraints: 
\begin{align}
I(F;S_{21},S_{31})= I(F; S_{12},S_{32})=I(F;S_{13},S_{23})=0,
\end{align}
to indicate the secure repair of nodes $1,2$ and $3$ respectively. If the eavesdropper can read $(S_{21}, S_{31})$, i.e., the repair data for $W_{1}$, then two cases can arise:
\begin{itemize}
\item if there is no redundancy in the repair process ($\alpha=d\beta$), then it is clear that if the stored data is secure, then the repair data will also be secure.

\item However, for $\alpha< d\beta$, more information about the file could be leaked in general. This  aspect is not captured by the proof for the Type-I attack and one needs to carefully deal with the security of the repair process.
\end{itemize}


We have the following sequence of bounds for secure repair of node $1$,
\begin{align}
H(F)&=H(F|S_{21},S_{31})\\
&=H(F,S_{21},S_{31})-H(S_{21},S_{31})\\
&\leq H(F,W_2,S_{21},S_{31})-H(S_{21},S_{31})\\
&=H(W_2,S_{21},S_{31}) + \underbrace{H(F|W_{2}, S_{21}, S_{31})}_{=0 \text{ (file regeneration)}}-H(S_{21},S_{31})\label{U-regen}\\
&=H(W_2,S_{21},S_{31})-H(S_{21},S_{31})\\
&=H(W_2,S_{21},S_{31})-H(S_{21},S_{31}, W_{1}) + \underbrace{H(W_{1}|S_{21}, S_{31})}_{=0 \text{ (repair of node 1)}}\\
&=H(W_2,S_{21},S_{31})-H(S_{21},S_{31}, W_{1})\\
&=H(W_2,S_{21},S_{31})-H(S_{21},S_{31}, W_{1}, S_{12}, S_{13}) + \underbrace{H(S_{12}, S_{13}|W_{1}, S_{21}, S_{31})}_{=0} \label{U-func}\\
&\leq H(W_2,S_{21},S_{31})-H(S_{21},S_{31},S_{12},S_{13}),\label{node_1_full}
\end{align}
where \eqref{U-regen} follows from the file regeneration requirement from nodes $(1, 2)$, \eqref{U-func} follows from the fact that $(S_{12}, S_{13})$ are functions of $W_{1}$. Similarly, for secure repair of node $2$, we have
\begin{align}
H(F)&\leq H(W_1,S_{12},S_{32})-H(S_{12},S_{32},S_{21},S_{23}).\label{node_2_full}
\end{align}

Adding \eqref{node_1_full} and \eqref{node_2_full}, we have
\begin{align}
2H(F)&\leq H(W_2,S_{21},S_{31})+H(W_1,S_{12},S_{32})-H(S_{21},S_{31},S_{12},S_{13})-H(S_{12},S_{32},S_{21},S_{23}) \\
&=H(W_2,S_{21},S_{31})+H(W_1,S_{12},S_{32})-2H(S_{21},S_{12})-H(S_{31},S_{13}|S_{21},S_{12})-H(S_{23},S_{32}|S_{21},S_{12})\\
&\leq H(W_2,S_{21},S_{31})+H(W_1,S_{12},S_{32})-H(S_{21},S_{12})-H(S_{12},S_{21},S_{13},S_{31},S_{23},S_{32})\\
&= H(W_2,S_{21},S_{31})+H(W_1,S_{12},S_{32})-H(S_{21},S_{12})-H(S_{12},S_{21},S_{13},S_{31},S_{23},S_{32}, W_{2})\label{reducing_the_bound}\\
&\leq H(W_2,S_{21},S_{31})+H(W_1,S_{12},S_{32})-H(S_{21},S_{12})-H(W_2,S_{21},S_{31})\label{reducing_the_bound-2} \\
&=H(W_1,S_{12},S_{32})-H(S_{21},S_{12})\\
&\leq H(W_1,W_3,S_{12},S_{32})-H(S_{21},S_{12}) \\
&= H(W_{3}, S_{12}) + \underbrace{H(W_{1}, S_{32}|W_{3}, S_{12})}_{=0 } - H(S_{21}, S_{12})\\
&= H(W_{3}, S_{12}) - H(S_{21}, S_{12})\label{repair_cause}\\
&\leq H(W_3, S_{12})-H(S_{12})\\
&\leq H(W_3) + H(S_{12})-H(S_{12})\\
&= H(W_{3})\\
&\leq \alpha.\label{FINAL-322}
\end{align}
where 
\begin{itemize}
\item \eqref{reducing_the_bound} is obtained by noting that $W_{2}$ is a function of the repair data $S_{12},S_{32}$,
\item \eqref{reducing_the_bound-2} follows from the fact that conditioning reduces entropy,
\item \eqref{repair_cause} is obtained by noting the fact that $(W_{1}, W_{2})$ can be obtained from $(W_3,S_{12})$. To note this, first observe that $(S_{32}, S_{31})$ are functions of $W_{3}$, hence using $(S_{32}, S_{12})$, we can repair $W_{2}$ (and subsequently obtain $S_{21}$). Thus, using $(S_{21}, S_{31})$, we can repair $W_{1}$.  
\end{itemize}
Hence from \eqref{FINAL-322}, we have $H(F)\leq \frac{\alpha}{2}$. Thus along with the bound $H(F)\leq \beta$, we have the proof for
\begin{align}
H(F)&\leq \min\left(\frac{\alpha}{2}, \beta\right)
\end{align}
Interestingly, this tradeoff between $(\alpha, \beta)$ only has one efficient point (see Fig.~\ref{fig:Theorem322}), corresponding to the case when $\alpha=2\beta$ (or the MBR point for the $(3,2,2)$ DSS).  
Extending this approach to more general $(n,k,d)$-DSS is far from trivial. The proof presented above essentially couples two security constraints corresponding to the secure repair of nodes $1$ and $2$. In general, we believe that it suffices to consider the coupling between $\binom{k}{l}$ security constraints, (each constraint corresponding to the secure repair of $l$ nodes). As we show in the Appendix, this statement turns out to be true for the $(4,2,3)$, $(4,3,3)$ and the $(n, n-1, n-1)$ DSS and helps us in establishing the corresponding optimal tradeoff regions.

%


\section{Achievability Proofs}\label{Achievability}
In this section, we present coding schemes that achieve the bounds mentioned in Theorems~\ref{Theorem322}-\ref{Theoremnnn}. We first present the achievability of the MBR point under Type-II secrecy constraint. Later, the coding schemes that achieve a various spectrum of points beyond the MBR point in the presence of a Type-I adversary are presented. 
\subsection{Achievability under Type-II security constraint}
As mentioned earlier, the MBR point is the only valid point in the $(\alpha,\beta)$ tradeoff region in the presence of Type-II adversary. The MBR point is defined by the $(\alpha,\beta)$ relationship 
$\alpha=d\beta$\cite{Dimakis_Intro}. Substituting this in \eqref{eq:functional_repair_tradeoff_secure}, we get that the optimal secure file size $\mathcal{B}^S$ must satisfy
\begin{align}\label{eq:MBRBound}
\mathcal{B}^S\leq \left(kd-\begin{pmatrix}k\\2\end{pmatrix}\right)\beta-\left(ld-\begin{pmatrix}l\\2\end{pmatrix}\right)\beta.
\end{align}
Notice that \eqref{eq:MBRBound} is identical to the tradeoff regions specified by Theorems~\ref{Theorem322}-\ref{Theoremnnn} with the corresponding values of $l$. This equivalence is further explained below. 

Secure codes that achieve this MBR point for a general $(n,k,d)$-DSS with any $l<k$ compromised nodes have been described in \cite{Shah_Rashmi_Secure} for both Type-I and Type-II attacks (since $\alpha=d\beta$ at the MBR point, the eavesdropper does not get any additional information by wiretapping the repair process i.e., Type-I and Type-II attacks are equivalent at the MBR point). 
For example, in the $(3,2,2)$-DSS with $l=1$, the MBR point in \eqref{eq:MBRBound} is simplified as $\mathcal{B}^S_{II}=\beta; \alpha=2\beta$, which is the $\alpha=2,\beta=1$, $\mathcal{B}^S_{II}=1$ point shown in Fig.~\ref{fig:Theorem322}, whose achievability is given in Fig.~\ref{fig:Intro2}. In the $(4,3,3)$-DSS with $l=1$, \eqref{eq:MBRBound} is given by $\mathcal{B}^S_{II}=3\beta; \alpha=3\beta$, which is also the $\alpha=3,\beta=1$, $\mathcal{B}^S_{II}=3$ point shown in Fig.~\ref{fig:Theorem433}. Along similar lines, with $l=2$, the MBR point in \eqref{eq:MBRBound} is given by $\mathcal{B}^S_{II}=\beta; \alpha=3\beta$, i.e., the $\alpha=3,\beta=1$, $\mathcal{B}^S_{II}=1$ point shown in Fig.~\ref{fig:Theorem433L2}. For the $(4,2,3)$-DSS, with $l=1$, we have $\alpha=3$, $\beta=1$ and $\mathcal{B}^S_{II}=2$, i.e., MBR point in Fig.~\ref{fig:Theorem423L1}. For the $(n,n-1,n-1)$-DSS, with $l=n-2$, we have $\alpha=n-1, \beta=1$ and $\mathcal{B}^S_{II}=1$, i.e., the MBR point in Fig.~\ref{fig:TheoremnnnLn2}. For more information on the codes that achieve the MBR point, please see \cite{Shah_Rashmi_Secure}.

\subsection{Achievability under Type-I security constraint}
At the MBR point, the Type-I and Type-II security constraints are equivalent and hence the MBR point is achievable under both these constraints \cite{Shah_Rashmi_Secure}. Beyond this MBR point, a spectrum of points in the secure storage-vs-exact-repair-bandwidth tradeoff region are achievable in the presence of a Type-I adversary. We next present the various coding schemes that achieve these points. 
\subsubsection{$(\alpha,\beta)=(1,1)$, $\mathcal{B}^S_I=1$ in a $(3,2,2)$-DSS, $l=1$} \label{Scheme:322L1a1b1B1}
       
The coding scheme that achieves this point is shown in Fig.~\ref{fig:Intro}. In this scheme, $X\in \mathbb{F}_q$ and $K$ the secure key, is a random variable that is uniformly distributed over $\mathbb{F}_q$ such that $q>3$. From Fig.~\ref{fig:Intro}, it is easy to see that the repair and the reconstruction processes are straightforward.  

\subsubsection{$(\alpha,\beta)=(2,1)$, $\mathcal{B}^S_I=2$ in a $(4,2,3)$-DSS, $l=1$} \label{Scheme:423L1a2b1B2}
This point corresponds to the normalized $(\alpha,\beta)$ pair i.e., 
$(\bar{\alpha},\bar{\beta})=(1,\frac{1}{2})$ in the Fig.~\ref{fig:Theorem423L1}. Let $a_1,a_2$ denote the message symbols to be stored in the DSS. Choose keys $k_1$ and $k_2$ to securely store the data on the DSS. Here $a_1,a_2,k_1,k_2 \in \mathbb{F}_5$. Let $x_1,x_2,x_3,x_4$ denote $4$ linearly independent combinations of the message and key symbols. They are stored in the DSS as shown in Table~\ref{tab:Theorem423CodeL1A2B1}. Notice that the eavesdropper cannot infer any information by observing the data contents on any single storage node. Further, the repair and reconstruction processes are straightforward. 

\begin{table}[h]
\centering
\caption{A $(4,2,3)$ secure exact repair code for $\left(\alpha,\beta\right)=\left(2,1\right)$, $\mathcal{B}^S=2$. }
\begin{tabular}{|c|c|c|c|}\hline
&first symbol&second symbol \\ \hline
node $1$ &$x_1$&$x_2$ \\ \hline 
node $2$ &$x_3$&$x_4$  \\ \hline
node $3$ &$x_1+x_3$&$x_2+x_4$ \\ \hline
node $4$ &$x_1+x_4$&$x_2+x_3$  \\ \hline
\end{tabular}
\vspace{-5pt}
\label{tab:Theorem423CodeL1A2B1}
\end{table}

\subsubsection{$(\alpha,\beta)=(1,1)$, $\mathcal{B}^S_I=2$ in a $(4,3,3)$-DSS, $l=1$} \label{Scheme:433L1a1b1B1}
This point corresponds to the normalized $(\alpha,\beta)$ pair i.e., 
$(\bar{\alpha},\bar{\beta})=(\frac{1}{2},\frac{1}{2})$ in the Fig.~\ref{fig:Theorem433}. 
Below, we present a coding scheme that can store a file of size $\mathcal{B}^S=2$ securely in a $(4,3,3)$-DSS with $l=1$ when each node stores $\alpha=1$ unit of data and sends $\beta=1$ unit of data for the repair of failed nodes. \begin{table}[h]
\centering
\caption{A $(4,3,3)$ secure exact repair code for $\left(\alpha,\beta\right)=\left(1,1\right)$, $\mathcal{B}^S=2$. }
\begin{tabular}{|c|c|c|c|}\hline
node $1$&node $2$& node $3$& node $4$ \\ \hline
$a_1+k$&$a_1+k$&$a_1+a_2+k$&$k$ \\ \hline
\end{tabular}
\label{tab:Theorem433CodeL1A1B1}
\end{table}

Let $a_1,a_2\in \mathbb{F}_2$ denote the message symbols to be stored in the DSS. Choose a key $k$ that is uniformly distributed over $\mathbb{F}_2$ in order to securely store the data on the DSS. The resulting DSS is shown in Table.~\ref{tab:Theorem433CodeL1A1B1}. Since the eavesdropper is unaware of the key $k$, it cannot decode either $a_1$ or $a_2$ from any of the nodes in the DSS, thereby securing the DSS against Type-I attacks. \

Further, it is easy to see that the file $(a_1,a_2)$ can be recovered from any $k=3$ nodes (by using $3$ linearly independent combinations of $3$ symbols $a_1,a_2$ and $k$). Consider the repair process of the $1$st node. The nodes $2,3$ and $4$ send their data contents in order to repair this failed node. See that $a_1$ and $k$ can be recovered from these three data symbols and that $a_1+k$, the contents of the $1$st node can be restored exactly.

\subsubsection{$(\alpha,\beta)=(3,2)$, $\mathcal{B}^S_I=5$ in a $(4,3,3)$-DSS, $l=1$} \label{Scheme:433L1a3b2B5}
This point corresponds to the normalized $(\alpha,\beta)$ pair i.e., 
$(\bar{\alpha},\bar{\beta})=(\frac{3}{5},\frac{2}{5})$ in the Fig.~\ref{fig:Theorem433}. 
This can be shown through a coding scheme that can securely store a file of size $\mathcal{B}^S=5$ over $n=4$ nodes where each node stores $3$ units of data and sends $2$ units of data for repair of other nodes. Let $\left(a_1,a_2,a_3,a_4,a_5\right)\in\mathbb{F}_q$ denote the message symbols to be stored in the DSS. Let $\left(k_1,k_2,k_3\right)$ be the secure keys that are used by the DSS to keep the data secure from the eavesdropper. It is assumed that these keys are uniformly distributed over $\mathbb{F}_q$ for some $q>8$. 

Let $A=[a_1,a_2,a_3,a_4,a_5]^T$ and $K=[k_1,k_2,k_3]^T$ represent the message and the secure key vectors. Let 
\begin{align}
X=[A\ K]\begin{bmatrix}G_A\\ G_K\end{bmatrix}=G_AA+G_KK,
\end{align}
be the $1\times 12$ vector of symbols to be stored on a DSS. Let $\begin{bmatrix}G_A\\ G_K\end{bmatrix}$ be a $8\times 12$ Vandermonde matrix such that the $i$th row is defined as $[1,i,i^2,i^3,\ldots,i^{11}]$ where $i=1,2,\ldots,8$. 
Here, $G_A$ is a $5\times 12$ sub matrix obtained from the above Vandermonde matrix given by
\begin{align}
\begin{bmatrix}1,1,1^2,\ldots,1^{11}\\ 
1,2,2^2,\ldots,2^{11}\\
\vdots\\
1,5,5^2,\ldots,5^{11}\\
\end{bmatrix}
\end{align}
and $G_K$ is a $3\times 12$ sub matrix given by 
\begin{align}
\begin{bmatrix}
1,6,6^2,\ldots,6^{11}\\
1,7,7^2,\ldots,7^{11}\\
1,8,8^2,\ldots,8^{11}\\
\end{bmatrix}
\end{align} 
Since we use a $8\times 12$ Vandermonde matrix, we define all the elements over $\mathbb{F}_9$. The resulting elements of $X$ are stored in the DSS as shown in Table~\ref{tab:Theorem433Code}.

In order to achieve secrecy, i.e., the eavesdropper should not get any information about $\{a_j\}_{j=1}^5$ by observing the contents of any single node, the following condition must hold true., 
\begin{align}
I(A;[X]_3)=0
\end{align}
where $[X]_3$ is any $3$ elements of the vector $X$ (since any $3$ elements of $X$ are linearly independent combinations of $A$ and $K$). We can represent $[X]_3$ as 
\begin{align}
[X]_3=[A\ K]\begin{bmatrix}\hat{G}_A\\ \hat{G}_K\end{bmatrix}_{8\times 3},
\end{align}
where $\hat{G}_A$, $\hat{G}_K$ are sub matrices of $\begin{bmatrix}G_A\\ G_K\end{bmatrix}$ that correspond to $[X]_3$. Notice that these sub matrices are also Vandermonde matrices. Thus we have the following set of inequalities,
\begin{align}
I(A;[X]_3)
&=I(A;\hat{G}_AA+\hat{G}_KK) \nonumber \\
&=H(\hat{G}_AA+\hat{G}_KK)-H(\hat{G}_AA+\hat{G}_KK|A) \nonumber \\
&=H(\hat{G}_AA+\hat{G}_KK)-H(\hat{G}_KK) \nonumber \\
&\leq 3-\mathsf{rank}\left(\hat{G}_K\right)\nonumber\\
&= 3-3=0\nonumber,
\end{align}
The last two inequalities follow from the fact that the sub-Vandermonde matrices have a maximum rank of $3$.

\begin{table}[t]
\centering
\caption{A $(4,3,3)$ secure exact repair code for $\left(\alpha,\beta\right)=\left(3,2\right)$, $\mathcal{B}^S=5$. }
\begin{tabular}{|c|c|c|c|}\hline
&first symbol&second symbol& third symbol \\ \hline
node $1$ &$x_1$&$x_2$&$x_{9}=x_6+x_7$ \\ \hline 
node $2$ &$x_3$&$x_4$&$x_{10}=x_1+x_8$  \\ \hline
node $3$ &$x_5$&$x_6$&$x_{11}=x_2+x_3$ \\ \hline
node $4$ &$x_7$&$x_8$&$x_{12}=x_4+x_5$  \\ \hline
\end{tabular}
\vspace{-5pt}
\label{tab:Theorem433Code}
\end{table}
\begin{table}[t]
\centering
\caption{Repair procedure of a $(4,3,3)$ secure exact repair code when node $1$ fails. }
\begin{tabular}{|c|c|c|c|}\hline
&first symbol&second symbol \\ \hline
node $2$&$x_3$&$x_{10}=x_1+x_8$  \\ \hline
node $3$&$x_6$&$x_{11}=x_2+x_3$ \\ \hline
node $4$&$x_7$&$x_8$  \\ \hline
\end{tabular}
\vspace{-4pt}
\label{tab:Theorem433Repair}
\end{table}

\emph{Repair Process}: Since this is a cyclic construction, every node is equivalent. Without loss of generality, it suffices to consider the repair of any single node. Consider the case when node $1$ fails and the other nodes participate in the repair process. Since $\beta=2$, each node sends $2$ symbols to aid the repair of the $1$st node. The repair process of node $1$ is shown in Table~\ref{tab:Theorem433Repair}. The first symbol of node $1$ is recovered by combining the second symbols sent by nodes $2$ and $4$. The second symbol of node $1$ is obtained by a combination of the first symbol sent by node $2$ and the second symbol sent by node $3$. Finally, the third symbol of node $1$ is recovered by a combination of the first symbols sent by nodes $3$ and $4$. Hence, upon receiving the $6$ symbols, it can be seen that the $1$st node can exactly recover its original symbols, thereby satisfying the requirements of an exact repair code.  

\begin{remark}
{\em This code construction can be easily extended to the exact repair case in the absence of an eavesdropper. In the absence of an eavesdropper (no requirement for keys), this code can store a file size of $8$ $\left(\{x_j\}_{j=1}^8, \textnormal{no linear combinations required} \right)$ in $n=4$ nodes with each node storing $\alpha=3$ units of data and sending $\beta=2$ units for the repair process. Thus this code achieves a normalized storage-bandwidth tradeoff point $(\frac{3}{8},\frac{1}{4})$ which is shown to be optimal in \cite{Chao_Tian}. Further, this code is also efficient in terms of the disk reads required for a repair process compared to the code proposed for the exact repair of a $(4,3,3)$-DSS in \cite{Chao_Tian}. It is seen that the number of disk reads required in this proposed code is $6$ ($6$ symbols obtained directly from the nodes) while it is $9$ in the case of the code proposed in \cite{Chao_Tian} (see \cite{Chao_Tian} for more details). This behavior is akin to the codes proposed in \cite{PiggyBacking} that are designed to be data-read and download efficient during a repair process. }
\end{remark}

\subsubsection{$(\alpha,\beta)=(1,1)$, $\mathcal{B}^S=1$ in a $(4,3,3)$-DSS, $l=2$}
Let $a\in \mathbb{F}_q$ denote the message symbol to be stored on the DSS. Since $l=2$, any two nodes need to be secure from the eavesdropping attack. Hence we choose two keys $k_1, k_2$ that are uniformly distributed in $\mathbb{F}_q$ in order to securely store $a$. Let $X=[x_1,x_2,x_3,x_4]^T$ denote the storage contents of the $4$ nodes in the DSS such that
\begin{align}
X=[G_A\ G_K]\begin{bmatrix}a\\ k_1\\ k_2\end{bmatrix}=G_Aa+G_K\begin{bmatrix}k_1\\ k_2\end{bmatrix},
\end{align}
where $G_A$ and $G_K$ are chosen such that $\mathsf{rank}\left([G_A\ G_K]\right)=3$ and $\mathsf{rank}\left(G_K\right)=2$ (for instance Vandermonde matrices similar to Section~\ref{Scheme:433L1a3b2B5}). 

$\{x_i\}_{i=1}^4$ are the linear combinations of the message symbol $a$ and the secure keys $k_1,k_2$. The rank of the concatenated matrix $[G_A\ G_K]$ ensures that the file can be recovered from any $3$ out of the $4$ nodes in the DSS ($3$ linearly independent combinations of $3$ symbols), thereby satisfying the reconstruction property of the DSS. By following the similar  arguments in Section~\ref{Scheme:433L1a3b2B5}, it can be shown that the rank of the matrix $G_K$ which is $2$, ensures that the storage contents on any two nodes are secure from a Type-I attack by an eavesdropper. 

\begin{remark}
{\em The above coding scheme can be extended to a general $(n,n-1,n-1)$-DSS with $l=n-2$ to achieve the $(\alpha,\beta)=(1,1)$, $\mathcal{B^S}=1$ point under exact repair and Type-I secrecy as shown in Fig.~\ref{fig:TheoremnnnLn2}.}
\end{remark}

\begin{remark}
{\em Although the same $(\alpha,\beta)=(3,1)$ pair is achievable under Type-II secrecy when $l=1$ or $l=2$ for the $(4,3,3)$-DSS, it is seen that the maximum file size that can be securely stored when $l=1$ is $\mathcal{B}^S=3$ while it is $\mathcal{B}^S=1$ when $l=2$. Intuitively, when $2$ nodes in the DSS are compromised, the secrecy constraints are stringent compared to the case when only one node is compromised and hence the file size that can be securely stored decreases when the number of compromised nodes increases.}
\end{remark}

\section{Conclusion}\label{Conclusion}
Securing distributed storage systems against passive eavesdropping attacks is addressed in this paper. A complete characterization of the storage-bandwidth tradeoff region is provided for a $(n,k,d)$-DSS for $n\leq 4$ and $l<k$ under exact repair and Type-I, Type-II secrecy constraints. These results have been extended to the $(n,n-1,n-1)$-DSS when $n-2$ nodes are compromised. Novel regenerating codes that are read and download repair efficient are presented in order to achieve these optimal secure tradeoff regions. The converse proof for the optimal tradeoff region under Type-I secrecy constraint was leveraged off the proof presented for the exact repair tradeoff region in the absence of an eavesdropper. Novel converse proofs were presented under the Type-II secrecy constraints, thereby characterizing the maximum file size that can be securely stored in a DSS. As expected, the file size that can be securely stored decreases when the number of compromised nodes increases in the DSS. The gap in the file size that can be securely stored under Type-I and Type-II attacks increases as $n$ increases, thereby indicating the severe limitations of the DSS under Type-II attacks. Further, the secure storage-vs-exact repair bandwidth tradeoff region obtained for the Type-II attacks indicates that the MBR point is the only achievable point in the corresponding tradeoff region. Extending these results to a general $(n,k,d)$-DSS is part of our ongoing work. 

\section*{Acknowledgement}
This work was supported in part by L3 Communications. The authors would also like to thank Dr. Soheil Mohajer for helpful discussions regarding the converse proof of $(4,2,3)$-DSS.

\section{Appendix}
\subsection{Proof of Theorem \ref{Theorem4232}: $(4,2,3)$-DSS, $l=1$}\label{ConvProofTheorem4232}
In this section, we present the proof for the Type-II setting for the the $(4,2,3)$-DSS and $l=1$. In particular we will show that
\begin{align}
\mathcal{B}^{S}_{II}&\leq \min\left(\frac{2\alpha}{3},2\beta\right).
\end{align}
To this end, we focus on proving that  $\mathcal{B}^{S}_{II}\leq \frac{2\alpha}{3}$.
\begin{itemize}
\item File regeneration from any $k=2$ nodes:
\begin{align}
H(F|W_{1}, W_{2})&= 0\label{Regen12}\\
H(F|W_{1}, W_{3})&= 0\label{Regen13}\\
H(F|W_{1}, W_{4})&= 0\label{Regen14}\\
H(F|W_{2}, W_{3})&=0\label{Regen23}\\
H(F|W_{2}, W_{4})&=0\label{Regen24}\\
H(F|W_{3}, W_{4})&=0\label{Regen34}.
\end{align}
\item Exact repair requirements:
\begin{align}
H(W_{1}|S_{21}, S_{31}, S_{41})&=0\label{W1repair4}\\
H(W_{2}|S_{12}, S_{32}, S_{42})&=0\label{W2repair4}\\
H(W_{3}|S_{13}, S_{23}, S_{43})&=0\label{W3repair4}\\
H(W_{4}|S_{14}, S_{24}, S_{34})&=0\label{W4repair4}.
\end{align}
\item Secure repair of any $l=1$ node:
\begin{align}
I(F; S_{21}, S_{31}, S_{41})&=0 \label{R1secure4}\\
I(F; S_{12}, S_{32}, S_{42})&=0  \label{R2secure4}\\
I(F; S_{13}, S_{23}, S_{43})&=0  \label{R3secure4}\\
I(F; S_{14}, S_{24}, S_{34})&=0  \label{R4secure4}.
\end{align}
\item Repair data from a node is a function of stored data:
\begin{align}
H(S_{12}, S_{13},S_{14}|W_{1})&= 0\label{4Func1}\\
H(S_{21}, S_{23}, S_{24}|W_{2})&= 0\label{4Func2}\\
H(S_{31}, S_{32}, S_{34}|W_{3})&= 0\label{4Func3}\\
H(S_{41}, S_{42}, S_{43}|W_{4})&= 0\label{4Func4}.
\end{align}
\end{itemize}
For secure repair of node $1$, we have
\begin{align}\label{node1_423}
H(F)&=H(F|S_{21},S_{31},S_{41})\\
&=H(F,S_{21},S_{31},S_{41})-H(S_{21},S_{31},S_{41})\\
&\stackrel{(\ref{W1repair4})}=H(F,S_{21},S_{31},S_{41})-H(S_{21},S_{31},S_{41},S_{12})\\
&=H(F,S_{21},S_{31},S_{41})-H(S_{12},S_{21})-H(S_{31},S_{41}|S_{12},S_{21}).
\end{align}
Similarly, for secure repair of node $2$, we have
\begin{align}\label{node2_423}
H(F)&=H(F|S_{12},S_{32},S_{42})\\
&=H(F,S_{12},S_{32},S_{42})-H(S_{12},S_{32},S_{42})\\
&\stackrel{(\ref{W2repair4})}=H(F,S_{12},S_{32},S_{42})-H(S_{12},S_{31},S_{41},S_{21})\\
&=H(F,S_{12},S_{32},S_{42})-H(S_{12},S_{21})-H(S_{32},S_{42}|S_{12},S_{21}).
\end{align}

Adding \eqref{node1_423} and \eqref{node2_423}, we obtain
\begin{align}\label{sum_term}
2H(F)&=H(F,S_{21},S_{31},S_{41})+H(F,S_{12},S_{32},S_{42})-2H(S_{12},S_{21})-H(S_{31},S_{41}|S_{12},S_{21})-H(S_{32},S_{42}|S_{12},S_{21})\\
&\leq H(F,S_{21},S_{31},S_{41})+H(F,S_{12},S_{32},S_{42})-H(S_{12},S_{21})-H(S_{31},S_{41},S_{32},S_{42},S_{12},S_{21})\\
&\leq H(F,S_{21},S_{31},S_{41})+H(F,S_{12},S_{32},S_{42})-H(S_{21})-H(S_{31},S_{41},S_{32},S_{42},S_{12},S_{21}).
\end{align}

Notice that,
\begin{align}\label{combined_term}
H(S_{31},S_{41},S_{32},S_{42},S_{12},S_{21})&=H(S_{31},S_{41},S_{32},S_{42},S_{12},S_{21},W_1,W_2,F)\\
&\geq H(F,S_{21},S_{31},S_{41})\\
&=H(S_{21},S_{31},S_{41})+H(F|S_{21},S_{31},S_{41})\\
&\stackrel{(\ref{R1secure4})}=H(S_{21},S_{31},S_{41})+H(F).
\end{align}

Using \eqref{combined_term} in \eqref{sum_term}, we get 
\begin{align}\label{final_term}
2H(F)\leq H(F,S_{21},S_{31},S_{41})+H(F,S_{12},S_{32},S_{42})-H(S_{21})-(H(S_{21},S_{31},S_{41})+H(F)).
\end{align}

We bound the first term in the above equation as follows
\begin{align}
H(F,S_{21},S_{31},S_{41})&\leq H(F,S_{21},S_{31},S_{41},W_4)\\
&\stackrel{(\ref{Regen14})}=H(S_{21},S_{31},S_{41},W_4) \\
&=H(W_4)+H(S_{41}|W_4)+H(S_{21},S_{31}|S_{41},W_4)\\
&\leq H(W_4)+H(S_{21},S_{31}|S_{41})\\
&=H(W_4)+H(S_{21},S_{31},S_{41})-H(S_{41})\\
&\leq \alpha+H(S_{21},S_{31},S_{41})-H(S_{41}).\label{box1}
\end{align}
By symmetry, we can show that 
\begin{align}
H(F,S_{12},S_{32},S_{42})=H(F,S_{21},S_{31},S_{41}).
\end{align}
Hence, we can bound
\begin{align}\label{box2}
H(F,S_{12},S_{32},S_{42})&\leq \alpha+H(S_{21},S_{31},S_{41})-H(S_{41}).
\end{align}
Substituting \eqref{box1} and \eqref{box2} in \eqref{final_term}, we have
\begin{align}
2H(F)&\leq 2\alpha+2H(S_{21},S_{31},S_{41})-2H(S_{41})-H(S_{21})-H(F)-H(S_{21},S_{31},S_{41})\\
3H(F)&\leq2\alpha+H(S_{21},S_{31},S_{41})-2H(S_{41})-H(S_{21}). \label{generic_term}
\end{align}
Similarly, we have 
\begin{align}
3H(F)&\leq2\alpha+H(S_{21},S_{31},S_{41})-2H(S_{21})-H(S_{31}) \\
3H(F)&\leq2\alpha+H(S_{21},S_{31},S_{41})-2H(S_{21})-H(S_{41}) \\
3H(F)&\leq2\alpha+H(S_{21},S_{31},S_{41})-2H(S_{31})-H(S_{21}) \\
3H(F)&\leq2\alpha+H(S_{21},S_{31},S_{41})-2H(S_{31})-H(S_{41}) \\
3H(F)&\leq2\alpha+H(S_{21},S_{31},S_{41})-2H(S_{41})-H(S_{31}) \\
\end{align}
By using the fact that 
\begin{align}
H(S_{21},S_{31},S_{41})\leq H(S_{21})+H(S_{31})+H(S_{41}),
\end{align}
and summing all the above $6$ inequalities, we have
\begin{align}
3H(F)\leq2\alpha\implies\mathcal{B}_{II}^S\leq \frac{2\alpha}{3}.
\end{align}

\subsection{Proof of Theorem \ref{Theorem4331}: $(4,3,3)$-DSS, $l=1$}

\subsubsection{Converse for Type-I secrecy constraint}\label{ConvProof43311}
We note that for $(4,3,3)$-DSS and $l=1$, under Type-I secrecy constraint, the optimal $(\alpha,\beta)$-tradeoff is given by
\begin{align}
\mathcal{B}^{S}_{I}&\leq \min\left(\min(\alpha,2\beta)+ \min(\alpha, \beta), \frac{\alpha+6\beta}{3}\right).
\end{align}
We note that the following bound 
\begin{align}
\mathcal{B}^{S}_{I}&\leq \min(\alpha,2\beta)+ \min(\alpha, \beta).
\end{align}
follows directly from (\ref{eq:functional_repair_tradeoff_secure}). Hence, in the remainder of this section, we will show that 
\begin{align}
\mathcal{B}^{S}_{I}&\leq \frac{\alpha+6\beta}{3}.
\end{align}
This proof of this bound closely follows a recent result in \cite{Chao_Tian}  which presents the optimal $(\alpha,\beta)$-tradeoff for the $(4,3,3)$-DSS in the absence of secrecy constraint.  The key difference is carefully incorporating the Type-I secrecy constraint as we will highlight during the proof. We proceed (similar to \cite{Chao_Tian}) as follows:
\begin{align}
&8H(W_{4})+ 4H(S_{31}, S_{21}) + 4H(S_{32})\nonumber\\
&\geq 4H(S_{31}, S_{21}, W_{4}) + 4 H(S_{32}, W_{4})\nonumber\\
&\geq 4H(S_{31}, S_{21}, W_{4}) + 4 H(S_{32}, W_{4}) - 2I(S_{21}; W_{3}|W_{4}) - 2I(W_{3}; W_{4}|S_{32}, S_{34}, S_{24})\\
&\stackrel{(a)}= 4H(S_{31}, S_{21}, W_{4}) + 2 H(W_{4}) + 2H(W_{3}, W_{4}, S_{21})- 2H(W_{3}, W_{4}) + 2H(S_{32}, S_{34}, S_{24})\nonumber\\
&\quad + 2H(W_{3}, W_{4}, S_{24}, S_{32}, S_{34}) - 2H(W_{4}, S_{32}, S_{34}, S_{24})\nonumber\\
&\stackrel{(b)}\geq 4H(S_{31}, S_{21}, W_{4}) + 2 H(W_{4}) + 2\Big[H(F)+ H(W_{4})\Big] - 2H(W_{3}, W_{4}) + 2H(S_{32}, S_{34}, S_{24})\nonumber\\
&\quad + 2H(W_{3}, W_{4}, S_{24}, S_{32}, S_{34}) - 2H(W_{4}, S_{32}, S_{34}, S_{24})\label{PF1},
\end{align}
where 
\begin{itemize}
\item $(a)$ follows by expanding all the mutual information expressions; then using 
the fact that $(S_{32}, S_{34})$ is a function of $W_{3}$; and finally using the symmetry (henceforth referred in short by $(sym)$), which implies that 
$H(S_{32}, W_4)= H(S_{24}, W_{3})= H(S_{21}, W_{4})$.
In general we have $H(\mathcal{S},\mathcal{W})=H(\pi(\mathcal{S}),\pi(\mathcal{W}))$ where $\mathcal{S}\subseteq \{S_{ij}\}_{i,j=1}^{4}$, $i\neq j$, $\mathcal{W}\subseteq \{W_j\}_{j=1}^4$ and $\pi$ is a permutation of the indices $i,j$ \cite{Chao_Tian}. 

\item $(b)$ is the key inequality that invokes the secrecy constraint; and can be proved as follows
\begin{align}
H(W_{3}, W_{4}, S_{21})
&=H(F, W_{3}, W_{4}, S_{21})\nonumber\\
&\geq H(F, W_{4})\nonumber\\
&=  H(F) + H(W_{4}) - I(F;W_{4})\nonumber\\
&=  H(F) + H(W_{4}).\nonumber 
\end{align}
in which the first equality follows from the fact that the file $F$ can be recovered from $(W_{3}, W_{4}, S_{21})$ and the last equality follows from the Type-I secrecy constraint, i.e., $I(F; W_{4})=0$. 
\end{itemize}
Next, we lower bound the following term that appears in (\ref{PF1}):
\begin{align}
H(S_{31}, S_{21}, W_{4})+ H(S_{32}, S_{34}, S_{24})
&\stackrel{(sym)}= H(S_{34}, S_{24}, W_{1}) + H(S_{32}, S_{34}, S_{24})\nonumber\\
&= H(W_{1}|S_{34}, S_{24})+ H(S_{32}|S_{34}, S_{24}) + 2H(S_{34}, S_{24})\nonumber\\
&\geq H(W_{1}, S_{32}|S_{34}, S_{24})+  2H(S_{34}, S_{24})\nonumber\\
&= H(W_{1}, S_{32}, S_{34}, S_{24})+  H(S_{34}, S_{24})\nonumber\\
&\stackrel{(rep)}= H(W_{1}, W_{2}, W_{4}) + H(S_{34}, S_{24})\nonumber\\
&= H(W_{1}, W_{2}, W_{4}, F) + H(S_{34}, S_{24})\nonumber\\
&\geq H(W_{4}, F)+H(S_{34}, S_{24})\nonumber\\
&= H(W_{4})+ H(F) +   H(S_{34}, S_{24}),\label{PF2}
\end{align}
where the fourth equality in obtained by noticing that the file $F$ can be recovered using 
$(W_1,S_{32},S_{34}, S_{24})$ and the fact that $S_{ij}$ is a function of 
$W_i$. This is shown below. 
\begin{align}
H(W_1,S_{32},S_{34},S_{24})&=H(W_1,S_{14},S_{32},S_{34},S_{24}) \nonumber \\
&=H(W_1,W_4,S_{32})\nonumber\\
&=H(W_1,S_{12},W_4,S_{42},S_{32}) \nonumber \\
&=H(W_1,W_2,W_4), \nonumber
\end{align}
This property is henceforth denoted by $(rep)$ or repair in the rest of the converse proof. 
The equation (\ref{PF2}) follows from Type-I secrecy constraint. 
Substituting (\ref{PF2}) in (\ref{PF1}), we have the following 
\begin{align}
&8H(W_{4})+ 4H(S_{31}, S_{21}) + 4H(S_{32})\nonumber\\
&\geq 2H(S_{31}, S_{21}, W_{4}) + 2H(W_{4}) + 4\Big[H(F)+ H(W_{4})\Big] - 2H(W_{3}, W_{4})+ 2H(W_{3}, W_{4}, S_{24})\nonumber\\
&\quad -2H(W_{4},S_{32}, S_{34}, S_{24})+ 2H(S_{34}, S_{24})\nonumber\\
&\stackrel{({sym})}= 2H(S_{31}, S_{21}, W_{4}) + 2H(W_{4}) + 4\Big[H(F)+ H(W_{4})\Big] - 2H(W_{3}, W_{4})+ 2H(W_{3}, W_{4}, S_{24})\nonumber\\
&\quad -2H(W_{4},S_{32}, S_{34}, S_{24})+ 2H(S_{31}, S_{21})\nonumber\\
&\geq 2H(S_{31}, S_{21}, W_{4}) + 2H(S_{31}, S_{21}, W_{4})  - 2H(W_{3}, W_{4})+ 2H(W_{3}, W_{4}, S_{24})\nonumber\\
&\quad -2H(W_{4},S_{32}, S_{34}, S_{24}) + 4\Big[H(F)+ H(W_{4})\Big]\nonumber\\
&= 4H(S_{31}, S_{21}, W_{4}) +  4\Big[H(F)+ H(W_{4})\Big]  - 2H(W_{3}, W_{4})+ 2H(W_{3}, W_{4}, S_{24}) -2H(W_{4},S_{32}, S_{34}, S_{24})\label{PF3}.
\end{align}
Next, we have the following inequality:
\begin{align}
H(W_{3}, W_{4}, S_{24})- H(W_{3}, W_{4})&= H(S_{24}|W_{3}, W_{4})\nonumber\\
&\geq H(S_{24}|S_{13}, W_{3}, W_{4})\nonumber\\
&= H(S_{24}, S_{13}, W_{3}, W_{4})- H(S_{13}, W_{3}, W_{4}).\label{PF4}
\end{align}
Using (\ref{PF4}) to further lower bound (\ref{PF3}), we get
\begin{align}
8H(W_{4})+ 4H(S_{31}, S_{21}) + 4H(S_{32})
&\geq 4H(S_{31}, S_{21}, W_{4}) +  4\Big[H(F)+ H(W_{4})\Big] \nonumber\\
&\quad - 2H(S_{13}, W_{3}, W_{4})+ 2H(W_{3}, W_{4}, S_{24}, S_{13})\nonumber\\
&\quad -2H(W_{4},S_{32}, S_{34}, S_{24})\nonumber\\
&\stackrel{(rep)}= 2H(S_{31}, S_{21}, W_{4}) +  2H(S_{31}, S_{21}, W_{4}, W_{1})\nonumber\\
&\quad - 2H(S_{13}, W_{3}, W_{4})+ 2H(W_{3}, W_{4}, S_{24}, S_{13})\nonumber\\
&\quad -2H(W_{4},S_{32}, S_{34}, S_{24})+ 4\Big[H(F)+ H(W_{4})\Big] \nonumber\\
&\stackrel{(sym)}= 2H(S_{31}, S_{21}, W_{4}) +  2H(S_{13}, S_{23}, W_{4}, W_{3})\nonumber\\
&\quad - 2H(S_{13}, W_{3}, W_{4})+ 2H(W_{3}, W_{4}, S_{24}, S_{13})\nonumber\\
&\quad -2H(W_{4},S_{32}, S_{34}, S_{24})+ 4\Big[H(F)+ H(W_{4})\Big] \label{PF5}.
\end{align}
Next, we have the following inequality
\begin{align}
H(S_{13}, S_{23}, W_{4}, W_{3})+ H(W_{3}, W_{4}, S_{24}, S_{13})
 - H(S_{13}, W_{3}, W_{4})
&= H( S_{23}| S_{13}, W_{3}, W_{4})+ H(S_{24}|S_{13}, W_{3}, W_{4})\nonumber\\
&\qquad + H(S_{13}, W_{3}, W_{4})\nonumber\\
&\geq H( S_{23}, S_{24}| S_{13}, W_{3}, W_{4})+ H(S_{13}, W_{3}, W_{4})\nonumber\\
&= H( S_{23}, S_{24}, S_{13}, W_{3}, W_{4})\label{PF6}.
\end{align}
Using (\ref{PF6}) in (\ref{PF5}), we obtain
\begin{align}
8H(W_{4})+ 4H(S_{31}, S_{21}) + 4H(S_{32})
&\geq2H(S_{31}, S_{21}, W_{4}) + 4\Big[H(F)+ H(W_{4})\Big]\nonumber\\
&\hspace{15pt}+ 2H( S_{23}, S_{24}, S_{13}, W_{3}, W_{4}) - 2H(W_{4},S_{32}, S_{34}, S_{24})\nonumber\\
&\geq2H(S_{31}, S_{21}, W_{4}) + 4\Big[H(F)+ H(W_{4})\Big]\nonumber\\
&\hspace{15pt}+ 2H( S_{23}, S_{24}, S_{13}, W_{3}, W_{4}) - 2H(W_{4},S_{32}, S_{34}, S_{24}, S_{14})\nonumber\\
&\stackrel{(c)}= 2H(S_{31}, S_{21}, W_{4}) + 4\Big[H(F)+ H(W_{4})\Big]\nonumber\\
&\hspace{15pt}+ 2H( S_{23}, S_{24}, S_{13}, W_{3}, W_{4}) - 2H(S_{32}, S_{34}, S_{24}, S_{14})\label{PF7},
\end{align}
where $(c)$ follows as $W_{4}$ is a function of $(S_{14}, S_{24}, S_{34})$.
Next, we have 
\begin{align}
2H( S_{23}, S_{24}, S_{13}, W_{3}, W_{4})
&\stackrel{(sym)}= H( S_{23}, S_{24}, S_{14}, W_{3}, W_{4})+ H( S_{34}, S_{32}, S_{14}, W_{2}, W_{4})\nonumber\\
&\stackrel{(rep)}= H( S_{23}, S_{24}, S_{14}, S_{32}, S_{34},W_{3}, W_{4})\nonumber\\
&\qquad + H( S_{34}, S_{32}, S_{14}, S_{24}, S_{23}, W_{2}, W_{4})\nonumber\\
&= H(W_{3}, W_{4}| S_{14}, S_{24}, S_{34}, S_{23}, S_{32})\nonumber\\
&\quad  + H(W_{2}, W_{4}| S_{14}, S_{24}, S_{34}, S_{23}, S_{32}) \nonumber\\
&\quad + 2H(S_{14}, S_{24}, S_{34}, S_{23}, S_{32})\nonumber\\
&\geq H(W_{2}, W_{3}, W_{4}, S_{14}, S_{24}, S_{34}, S_{23}, S_{32}) \nonumber\\
&\quad + H(S_{14}, S_{24}, S_{34}, S_{23}, S_{32})\nonumber\\
&\geq H(W_{2}, W_{3}, W_{4})  + H(S_{14}, S_{24}, S_{34}, S_{23}, S_{32})\nonumber\\
&= H(W_{2}, W_{3}, W_{4}, F) + H(S_{14}, S_{24}, S_{34}, S_{23}, S_{32})\nonumber\\
&\geq H(W_{4}, F)  + H(S_{14}, S_{24}, S_{34}, S_{23}, S_{32})\nonumber\\
&= \Big[H(W_{4})+ H(F)\Big] + H(S_{14}, S_{24}, S_{34}, S_{23}, S_{32}),\label{PF8}
\end{align}
where (\ref{PF8}) follows from the Type-I secrecy constraint. Using (\ref{PF8}) in (\ref{PF7}) gives
\begin{align}
8H(W_{4})+ 4H(S_{31}, S_{21}) + 4H(S_{32})
&\geq 2H(S_{31}, S_{21}, W_{4}) + 5\Big[H(F)+ H(W_{4})\Big]\nonumber\\
&\quad - 2H(S_{32}, S_{34}, S_{24}, S_{14}) + H(S_{14}, S_{24}, S_{34}, S_{23}, S_{32})\nonumber\\
&\stackrel{(sym)}=  5\Big[H(F)+ H(W_{4})\Big]\nonumber\\
&\quad + H(S_{31}, S_{21}, W_{4}) + H(S_{14}, S_{24}, W_{3})\nonumber\\
&\quad - 2H(S_{32}, S_{34}, S_{24}, S_{14}) + H(S_{14}, S_{24}, S_{34}, S_{23}, S_{32})\nonumber\\
&\stackrel{(rep)}=  5\Big[H(F)+ H(W_{4})\Big]\nonumber\\
&\quad + H(S_{31}, S_{21}, W_{4}) + H(S_{14}, S_{24}, S_{32}, S_{34},W_{3})\nonumber\\
&\quad - 2H(S_{32}, S_{34}, S_{24}, S_{14}) + H(S_{14}, S_{24}, S_{34}, S_{23}, S_{32})\nonumber\\
&=  5\Big[H(F)+ H(W_{4})\Big]\nonumber\\
&\quad + H(S_{31}, S_{21}, W_{4}) + H(W_{3}|S_{14}, S_{24}, S_{32}, S_{34})\nonumber\\
&\quad +H(S_{23}| S_{14}, S_{24}, S_{32}, S_{34})\nonumber\\
&\geq  5\Big[H(F)+ H(W_{4})\Big]\nonumber\\
&\hspace{2pt} + H(S_{31}, S_{21}, W_{4}) + H(W_{3}, S_{23}|S_{14}, S_{24}, S_{32}, S_{34}).\label{PF9}
\end{align}
Finally, using symmetry, we write $H(S_{31}, S_{21}, W_{4})$ as:
\begin{align}
H(S_{31}, S_{21}, W_{4})&\stackrel{(sym)}= H(S_{14}, S_{34}, W_{2})\nonumber\\
&\stackrel{(rep)}= H(W_{2},S_{14}, S_{34}, S_{23}, S_{24})\label{TestA1}.
\end{align}
On the other hand, we also have 
\begin{align}
H(W_{3}, S_{23}|S_{14}, S_{24}, S_{32}, S_{34})
&= H(W_{3}, S_{23},S_{14}, S_{24}, S_{32}, S_{34}) - H(S_{14}, S_{24}, S_{32}, S_{34})\nonumber\\
&\stackrel{(sym)}= H(W_{3}, S_{23},S_{14}, S_{24}, S_{32}, S_{34}) - H(S_{14}, S_{24}, S_{32}, S_{23})\nonumber\\
&= H(W_{3}, S_{32}|S_{14}, S_{34}, S_{23}, S_{24})\nonumber\\
&\geq H(W_{3}, S_{32}|W_{2}, S_{14}, S_{34}, S_{23}, S_{24})\label{TestA2}.
\end{align}
Using (\ref{TestA1}) and (\ref{TestA2}), we have from (\ref{PF9})
\begin{align}
8H(W_{4})+ 4H(S_{31}, S_{21}) + 4H(S_{32})
&\geq  5\Big[H(F)+ H(W_{4})\Big] + H(W_{2}, W_{3}, S_{32}, S_{14}, S_{24}, S_{34}, S_{23})\nonumber\\
&\stackrel{(rep)}=  5\Big[H(F)+ H(W_{4})\Big] + H(W_{2}, W_{3}, W_{4}, S_{32}, S_{14}, S_{24}, S_{34}, S_{23})\nonumber\\
&\geq  5\Big[H(F)+ H(W_{4})\Big] + H(W_{2}, W_{3}, W_{4})\nonumber\\
&= 5\Big[H(F)+ H(W_{4})\Big] + H(W_{2}, W_{3}, W_{4}, F)\nonumber\\
&\geq 5\Big[H(F)+ H(W_{4})\Big] + H(W_{4}, F)\nonumber\\
&= 5\Big[H(F)+ H(W_{4})\Big] + H(W_{4})+ H(F)\label{PF10}\\
&= 6\Big[H(F)+ H(W_{4})\Big]\label{PF11},
\end{align}
where (\ref{PF10}) follows from Type-I secrecy constraint.
Hence, from (\ref{PF11}), we have 
\begin{align}
6H(F)&\leq 2H(W_{4}) + 4H(S_{31}, S_{21})+ 4H(S_{32})\\
&\leq 2\alpha + 12\beta,
\end{align}
which implies that $3H(F)\leq \alpha + 6\beta$ and hence we have the proof of the bound:
\begin{align}
\mathcal{B}^{S}_{I}\leq \frac{\alpha+6\beta}{3}.
\end{align}

\subsubsection{Converse for Type-II Constraint}\label{ConvProof43312}
For this case, we will show that the existing bound (\ref{eq:functional_repair_tradeoff_secure})
$\mathcal{B}^{S}_{II}\leq \min(\alpha, 2\beta)+ \min(\alpha, \beta)$
is strictly loose and the secure storage-repair bandwidth tradeoff is given by
\begin{align}
\mathcal{B}^{S}_{II}&\leq \min(\alpha, 3\beta).
\end{align}
To this end, we focus on proving that $\mathcal{B}^{S}_{II}\leq \alpha$. 
\begin{itemize}
\item File regeneration from any $k=3$ nodes:
\begin{align}
H(F|W_{1}, W_{2},W_{3})&= 0\label{Regen123}\\
H(F|W_{1}, W_{2}, W_{4})&= 0\label{Regen124}\\
H(F|W_{1}, W_{3}, W_{4})&= 0\label{Regen134}\\
H(F|W_{2}, W_{3}, W_{4})&=0\label{Regen234}.
\end{align}
\item Exact repair requirements:
\begin{align}
H(W_{1}|S_{21}, S_{31}, S_{41})&=0\label{W1repair4}\\
H(W_{2}|S_{12}, S_{32}, S_{42})&=0\label{W2repair4}\\
H(W_{3}|S_{13}, S_{23}, S_{43})&=0\label{W3repair4}\\
H(W_{4}|S_{14}, S_{24}, S_{34})&=0\label{W4repair4}.
\end{align}
\item Secure repair of any $l=1$ node:
\begin{align}
I(F; S_{21}, S_{31}, S_{41})&=0 \label{R1secure4}\\
I(F; S_{12}, S_{32}, S_{42})&=0  \label{R2secure4}\\
I(F; S_{13}, S_{23}, S_{43})&=0  \label{R3secure4}\\
I(F; S_{14}, S_{24}, S_{34})&=0  \label{R4secure4}.
\end{align}
\item Repair data from a node is a function of stored data:
\begin{align}
H(S_{12}, S_{13},S_{14}|W_{1})&= 0\label{4Func1}\\
H(S_{21}, S_{23}, S_{24}|W_{2})&= 0\label{4Func2}\\
H(S_{31}, S_{32}, S_{34}|W_{3})&= 0\label{4Func3}\\
H(S_{41}, S_{42}, S_{43}|W_{4})&= 0\label{4Func4}.
\end{align}
\end{itemize}
For secure repair of node $1$, we have
\begin{align}
H(F)
&\stackrel{(\ref{R1secure4})}= H(F|S_{21}, S_{31}, S_{41})\nonumber\\
&= H(F, S_{21}, S_{31}, S_{41}) - H(S_{21}, S_{31}, S_{41})\nonumber\\
&= H(F, S_{21}, S_{31}, S_{41}) - H(S_{21})- H(S_{31}|S_{21})- H(S_{41}|S_{31}, S_{21})\label{LA1}.
\end{align}
Similarly, for secure repair of node $2$, we have
\begin{align}
H(F)&\stackrel{(\ref{R2secure4})}= H(F|S_{12}, S_{32}, S_{42})\nonumber\\
&\stackrel{(\ref{W2repair4})}= H(F|S_{12}, S_{32}, S_{42}, W_{2})\nonumber\\
&\stackrel{(\ref{4Func2})}= H(F|S_{12}, S_{32}, S_{42}, W_{2}, S_{21}, S_{23})\nonumber\\
&\leq H(F|S_{21}, S_{23}, S_{42})\nonumber\\
&= H(F, S_{21}, S_{23}, S_{42}) - H(S_{21}, S_{23}, S_{42})\nonumber\\
&= H(F, S_{21}, S_{23}, S_{42}) - H(S_{21})- H(S_{23}|S_{21})- H(S_{42}|S_{23}, S_{21})\label{LA2}.
\end{align}
Adding (\ref{LA1}) and (\ref{LA2}), we obtain
\begin{align}
2H(F)
&\leq H(F, S_{21}, S_{31}, S_{41})+ H(F, S_{21}, S_{23}, S_{42}) - 2H(S_{21})- H(S_{31}|S_{21})- H(S_{23}|S_{21})  \nonumber\\
&\quad- H(S_{41}|S_{31}, S_{21})- H(S_{42}|S_{23}, S_{21})\nonumber\\
&\leq H(F, S_{21}, S_{31}, S_{41})+ H(F, S_{21}, S_{23}, S_{42}) - 2H(S_{21})- H(S_{31}, S_{23}|S_{21}) \nonumber\\
&\quad- H(S_{41}|S_{31}, S_{21})- H(S_{42}|S_{23}, S_{21})\nonumber\\
&= H(F, S_{21}, S_{31}, S_{41})+ H(F, S_{21}, S_{23}, S_{42})- H(S_{21})- H(S_{21}, S_{31}, S_{23}) \nonumber\\
&\quad- H(S_{41}|S_{31}, S_{21})- H(S_{42}|S_{23}, S_{21})\nonumber\\
&= H(F, S_{21}, S_{31}, S_{41})+ H(F, S_{21}, S_{23}, S_{42})- H(S_{21}, S_{31}, S_{23})  \nonumber\\
&\quad - H(S_{21}) - H(S_{41}| S_{21})- H(S_{42}|S_{21}) +I(S_{41}; S_{31}|S_{21})+ I(S_{42};S_{23}|S_{21})\nonumber\\
&\leq H(F, S_{21}, S_{31}, S_{41})+ H(F, S_{21}, S_{23}, S_{42})- H(S_{21}, S_{31}, S_{23})\nonumber\\
&\quad - H(S_{21})  - H(S_{41},S_{42}|S_{21})+I(S_{41}; S_{31}|S_{21})+ I(S_{42};S_{23}|S_{21})\nonumber\\
&= H(F, S_{21}, S_{31}, S_{41})+ H(F, S_{21}, S_{23}, S_{42})\nonumber\\
&\quad - H(S_{21}, S_{23}, S_{31})  - H(S_{41},S_{42}, S_{21}) +I(S_{41}; S_{31}|S_{21})+ I(S_{42};S_{23}|S_{21})\nonumber\\
&= \Big[H(F, S_{21}, S_{31}, S_{41})+ H(F, S_{21}, S_{23}, S_{42})\Big]\nonumber\\
&\quad - 2H(S_{21}, S_{23}, S_{31}) +I(S_{41}; S_{31}|S_{21})+ I(S_{42};S_{23}|S_{21})\label{Symmetrymain},
\end{align}
where in (\ref{Symmetrymain}), we have replaced $H(S_{41}, S_{42}, S_{21})$ by $H(S_{31}, S_{23}, S_{21})$
by invoking the following symmetry property:
\begin{align}
\hspace{-10pt}H(S_{41}, S_{42}, S_{21})&= H(S_{31}, S_{32}, S_{21})= H(S_{31}, S_{23}, S_{21})\nonumber.
\end{align}
Next, consider the term $H(F, S_{21}, S_{31}, S_{41})$ appearing in (\ref{Symmetrymain}):
\begin{align}
H(F, S_{21}, S_{31}, S_{41})
&\leq H(F, W_{4}, S_{23}, S_{21}, S_{31}, S_{41})\nonumber\\
&= H(W_{4}, S_{23}, S_{21}, S_{31})+ H(F, S_{41}| W_{4}, S_{23}, S_{21}, S_{31})\nonumber\\
&= H(W_{4}, S_{23}, S_{21}, S_{31})\label{Inbet1}\\
&= H(W_{4}) + H(S_{23}, S_{21}, S_{31}) - I(W_{4}; S_{21}, S_{31}, S_{23})\nonumber\\
&\leq H(W_{4}) + H(S_{23}, S_{21}, S_{31}) - I(S_{41}; S_{21}, S_{31}, S_{23})\label{Inbet2}\\
&\leq H(W_{4}) + H(S_{23}, S_{21}, S_{31}) - I(S_{41}; S_{21}, S_{31})\nonumber\\
&\leq H(W_{4}) + H(S_{23}, S_{21}, S_{31}) - I(S_{41}; S_{31}| S_{21})\nonumber\\
&\leq \alpha + H(S_{23}, S_{21}, S_{31}) - I(S_{41}; S_{31}| S_{21})\label{LA3}.
\end{align}
where (\ref{Inbet1}) follows from the fact that $S_{41}$ is a function of $W_{4}$; and the fact that the file $F$ can be recovered from $W_{4}, S_{23}, S_{21}, S_{31}$.
Equation (\ref{Inbet2}) follows from the fact that $S_{41}$ is a function of $W_{4}$ and by using data processing inequality.
Similarly, for the  term $H(F, S_{21}, S_{23}, S_{42})$  in 
(\ref{Symmetrymain}):
\begin{align}
H(F, S_{21}, S_{23}, S_{42})
&\leq H(F, W_{4}, S_{21}, S_{23}, S_{42}, S_{31})\nonumber\\
&= H(W_{4}, S_{23}, S_{21}, S_{31})+ H(F, S_{42}| W_{4}, S_{23}, S_{21}, S_{31})\nonumber\\
&= H(W_{4}, S_{23}, S_{21}, S_{31})\nonumber\\
&= H(W_{4}) + H(S_{23}, S_{21}, S_{31}) - I(W_{4}; S_{21}, S_{31}, S_{23})\nonumber\\
&\leq H(W_{4}) + H(S_{23}, S_{21}, S_{31}) - I(S_{42}; S_{21}, S_{31}, S_{23})\nonumber\\
&\leq H(W_{4}) + H(S_{23}, S_{21}, S_{31}) - I(S_{42}; S_{21}, S_{23})\nonumber\\
&\leq H(W_{4}) + H(S_{23}, S_{21}, S_{31}) - I(S_{42}; S_{23}| S_{21})\nonumber\\
&\leq \alpha + H(S_{23}, S_{21}, S_{31}) - I(S_{42}; S_{23}| S_{21})\label{LA4}.
\end{align}
Adding (\ref{LA3}) and (\ref{LA4}), we arrive at
\begin{align}
\Big[H(F, S_{21}, S_{31}, S_{41})+ H(F, S_{21}, S_{23}, S_{42})\Big]
&\leq 2\alpha + 2H(S_{23}, S_{21}, S_{31}) - I(S_{41}; S_{31}| S_{21}) - I(S_{42}; S_{23}| S_{21}).\label{LA5}
\end{align}
Substituting (\ref{LA5}) in (\ref{Symmetrymain}), we obtain
\begin{align}
2H(F)&\leq 2\alpha + 2H(S_{23}, S_{21}, S_{31}) - I(S_{41}; S_{31}| S_{21}) - I(S_{42}; S_{23}| S_{21})\nonumber\\
&\quad - 2H(S_{21}, S_{23}, S_{31}) +I(S_{41}; S_{31}|S_{21})+ I(S_{42};S_{23}|S_{21})\\
&= 2\alpha,
\end{align}
which yields $H(F)\leq \alpha$ and hence we have the proof for
\begin{align}
\mathcal{B}^{S}_{II}&\leq \alpha.
\end{align}

\subsection{Proof of Theorem \ref{Theorem4332}: $(4,3,3)$-DSS, $l=2$}\label{ConvProofTheorem4332}
Here, we focus on the case of $(4,3,3)$-DSS and $l=2$. For this case, we will show that the secure storage-repair bandwidth tradeoff is given by
\begin{align}
\mathcal{B}^{S}_{II}&\leq \min\left(\frac{\alpha}{3}, \beta\right).
\end{align}
For secure repair of any $l=2$ nodes, we have the following $\binom{4}{2}=6$ constraints:
\begin{align}
I(F; S_{21}, S_{31}, S_{41}, S_{12}, S_{32}, S_{42})&=0 \label{R12secure}\\
I(F; S_{21}, S_{31}, S_{41}, S_{13}, S_{23}, S_{43})&=0 \label{R13secure}\\
I(F; S_{12}, S_{32}, S_{42}, S_{13}, S_{23}, S_{43})&=0 \label{R23secure}\\
I(F; S_{21}, S_{31}, S_{41}, S_{14}, S_{24}, S_{34})&=0 \label{R14secure}\\
I(F; S_{12}, S_{32}, S_{42}, S_{14}, S_{24}, S_{34})&=0 \label{R24secure}\\
I(F; S_{13}, S_{23}, S_{43}, S_{14}, S_{24}, S_{34})&=0 \label{R34secure}.
\end{align}
With these secure repair constraints, and exact repair requirements, we will now show that 
\begin{align}
\mathcal{B}^{S}_{II}\leq \frac{\alpha}{3}.
\end{align}
For the secure repair of nodes $1$ and $2$, we have
\begin{align}
H(F)&\stackrel{(\ref{R12secure})}=H(F|S_{21}, S_{31}, S_{41}, S_{12}, S_{32}, S_{42})\nonumber\\
&\hspace{5pt}= H(F, S_{21}, S_{31}, S_{41}, S_{12}, S_{32}, S_{42}) - H(S_{21}, S_{31}, S_{41}, S_{12}, S_{32}, S_{42})\label{First12}.
\end{align}
We first focus on the term $H(F, S_{21}, S_{31}, S_{41}, S_{12}, S_{32}, S_{42})$ and bound it as follows:
\begin{align}
H(F, S_{21}, S_{31}, S_{41}, S_{12}, S_{32}, S_{42})
&\leq H(F, W_{4}, S_{21}, S_{31}, S_{41}, S_{12}, S_{32}, S_{42})\nonumber\\ 
&= H(W_{4}, S_{21}, S_{31}, S_{41}, S_{12}, S_{32}, S_{42}) + H(F|W_{4}, S_{21}, S_{31}, S_{41}, S_{12}, S_{32}, S_{42})\nonumber\\
&= H(W_{4}, S_{21}, S_{31}, S_{41}, S_{12}, S_{32}, S_{42}) + H(F|W_{4}, W_{1}, W_{2}, S_{21}, S_{31}, S_{41}, S_{12}, S_{32}, S_{42})\label{Last1}\\
&= H(W_{4}, S_{21}, S_{31}, S_{41}, S_{12}, S_{32}, S_{42})\label{Last2}\\
&= H(W_{4}, S_{21}, S_{31}, S_{41}, S_{12}, S_{32}, S_{42}, S_{23}, S_{13})\label{Last3}\\
&= H(W_{4}, S_{21}, S_{31}, S_{12}, S_{32}, S_{23}, S_{13})\label{Last4}\\
&= H(W_{4}, U_{123})\label{Last5},
\end{align}
where (\ref{Last1}) follows from the fact that $W_{1}$ can be obtained from $(S_{21}, S_{31}, S_{41})$
and $W_{2}$ can be obtained from $(S_{12}, S_{32}, S_{42})$. Next, (\ref{Last2}) follows from (\ref{Regen124}). 
(\ref{Last3}) follows from the fact that $W_{2}$ (and hence $S_{23}$) is a function of $(S_{12}, S_{32}, S_{42})$. 
Similarly, $W_{1}$ (and hence $S_{13}$) is a function of $(S_{21}, S_{31}, S_{41})$.
Finally, (\ref{Last4}) follows from the fact that $(S_{41}, S_{42})$ are functions of $W_{4}$. In (\ref{Last5}), we have defined
\begin{align}
U_{123}&\triangleq (S_{21}, S_{12}, S_{31}, S_{13}, S_{32}, S_{23}).\label{DefU123}
\end{align}
Substituting (\ref{Last5}) in (\ref{First12}), we have
\begin{align}
H(F)&\leq H(W_{4}, U_{123}) - H(S_{21}, S_{31}, S_{41}, S_{12}, S_{32}, S_{42})\nonumber\\
&\stackrel{(a)}= H(W_{4}, U_{123})- H(S_{21}, S_{31}, S_{41}, S_{12}, S_{32}, S_{42}, S_{23}, S_{13})\nonumber\\
&= H(W_{4}, U_{123})- H(S_{21}, S_{12}, S_{31}, S_{13}, S_{32}, S_{23})- H(S_{41}, S_{42}| S_{21}, S_{12}, S_{31}, S_{13}, S_{32}, S_{23})\nonumber\\
&\stackrel{(b)}= H(W_{4}, U_{123}) - H(U_{123})  - H(S_{41}, S_{42}| U_{123})\nonumber
\end{align}
where $(a)$ follows from the fact that $W_{2}$ (and hence $S_{23}$) is a function of $(S_{12}, S_{32}, S_{42})$. 
Similarly, $W_{1}$ (and hence $S_{13}$) is a function of $(S_{21}, S_{31}, S_{41})$. In $(b)$, we have used the 
variable $U_{123}$ as defined in (\ref{DefU123}). 

In summary, for secure repair of nodes $1$ and $2$, we have
\begin{align}
H(F)&\leq H(W_{4}, U_{123}) - H(U_{123})  - H(S_{41}, S_{42}| U_{123})\label{Sec12}.
\end{align}
Similarly, for secure repair of nodes $1$ and $3$, we have 
\begin{align}
H(F)&\leq H(W_{4}, U_{123}) - H(U_{123})  - H(S_{41}, S_{43}| U_{123})\label{Sec13},
\end{align}
and for secure repair of nodes $2$ and $3$, we can obtain
\begin{align}
H(F)&\leq H(W_{4}, U_{123}) - H(U_{123})  - H(S_{42}, S_{43}| U_{123})\label{Sec23}.
\end{align}
Summing up (\ref{Sec12}), (\ref{Sec13}) and (\ref{Sec23}), we obtain
\begin{align}
3H(F)&\leq 3H(W_{4}, U_{123}) - 3H(U_{123})- H(S_{41}, S_{42}| U_{123}) -  H(S_{41}, S_{43}| U_{123}) - H(S_{42}, S_{43}| U_{123}).\label{PFT31}
\end{align}
We next note that for any four random variables $(X,Y,Z, U)$, the following inequality holds:
\begin{align}
H(X,Y|U)\hspace{-1pt}+\hspace{-1pt}H(X,Z|U)\hspace{-1pt}+\hspace{-1pt}H(Y,Z|U)\hspace{-1pt}\geq\hspace{-1pt}2H(X,Y,Z|U).\label{INEQmain}
\end{align}
This inequality can be proved as follows:
\begin{align}
H(X,Y|U)+H(X,Z|U)+H(Y,Z|U)
&= H(X,Y|U)+H(Z|U) + H(X|Z,U)+H(Y,Z|U)\nonumber\\ 
&\geq H(X,Y|U)+H(Z|X,Y,U) + H(X|Y,Z,U)+H(Y,Z|U)\nonumber\\ 
&= 2H(X,Y, Z|U) \nonumber.
\end{align}

Using the inequality (\ref{INEQmain}) by letting $X=S_{41}$, $Y=S_{42}$, $Z=S_{43}$ and $U=U_{123}$ in (\ref{PFT31}), we obtain
\begin{align}
3H(F)&\leq 3H(W_{4}, U_{123}) - 3H(U_{123}) - H(S_{41}, S_{42}| U_{123}) -  H(S_{41}, S_{43}| U_{123}) - H(S_{42}, S_{43}| U_{123})\nonumber\\
&\leq 3H(W_{4}, U_{123}) - 3H(U_{123}) -2H(S_{41}, S_{42}, S_{43}| U_{123}) \nonumber\\
&= 3H(W_{4}, U_{123}) - H(U_{123}) -2H(S_{41}, S_{42}, S_{43}, U_{123}) \nonumber\\
&= 3H(W_{4}, U_{123}) - H(U_{123}) -2H(W_{4},S_{41}, S_{42}, S_{43}, U_{123})  + 2H(W_{4}|S_{41}, S_{42}, S_{43}, U_{123})\nonumber\\
&= 3H(W_{4}, U_{123}) - H(U_{123}) -2H(W_{4},S_{41}, S_{42}, S_{43}, U_{123})\label{PFT32}\\
&\leq 3H(W_{4}, U_{123}) - H(U_{123}) -2H(W_{4}, U_{123})\nonumber\\
&= H(W_{4}, U_{123}) - H(U_{123})\nonumber\\
&\leq H(W_{4})+ H(U_{123}) - H(U_{123})\nonumber\\
&= H(W_{4})\nonumber\\
&\leq \alpha,\label{FINALLLL}
\end{align}
where (\ref{PFT32}) follows from the fact that $H(W_{4}|S_{41}, S_{42}, S_{43}, U_{123})=0$. 
Hence, from (\ref{FINALLLL}), we have shown that $3H(F)\leq \alpha$, and thus we have the proof for
\begin{align}
\mathcal{B}^{S}_{II}&\leq \frac{\alpha}{3}.
\end{align}

\subsection{Proof of Theorem \ref{Theoremnnn}: $(n,n-1,n-1)$-DSS, $l=(n-2)$.}\label{ConvProofTheoremn}
In this section, we present the proof for the Type-II setting for the more general $(n,n-1,n-1)$-DSS and $l=n-2$.
In particular, we will show that 
\begin{align}
\mathcal{B}^{S}_{II}&\leq \min\left(\frac{\alpha}{n-1},\beta\right).
\end{align}
To this end, we focus on proving that  $\mathcal{B}^{S}_{II}\leq \frac{\alpha}{n-1}$.
From Type-II security requirement we require:
\begin{align}
I(F;S_{\pi_{1}}, S_{\pi_{2}}, \ldots, S_{\pi_{n-2}})=0,
\end{align}
where $S_{\pi_{r}}$ is the repair data (from the remaining $d=(n-1)$ alive nodes) that is used to repair the $\pi_{r}$th node. Note that there are $\binom {n}{ n-2}= n(n-1)/2$ such constraints;
each corresponding to the secure repair of a set of $l=(n-2)$ nodes. 

Let us consider the first $k=(n-1)$ nodes, i.e., nodes $1,2,\ldots, n-1$. For secure repair of any $l=(n-2)$ out of these $(n-1)$ nodes, we have $\binom{k}{l}= \binom{n-1}{n-2}=(n-1)$ constraints.
Before describing these constraints, we note that the repair data (coming from the remaining $(n-1)$ alive nodes)  for node $i$ is given by: 
\begin{align}
S_{i}&= (S_{1i}, \ldots, S_{(i-1)i}, S_{(i+1)i}, \ldots, S_{n i}).\label{DefnSi}
\end{align}
Using this, we define
\begin{align}
S_{[1:n-1]}\triangleq (S_{1}, S_{2}, \ldots, S_{n-1}),
\end{align}
where $S_{[1:n-1]}$ is the collective repair data that is used to repair the first $k=(n-1)$ nodes. 
Next, we define 
\begin{align}
U_{ij}&\triangleq (S_{ij}, S_{ji}),
\end{align}
where $U_{ij}$ consists of the repair data $S_{ij}$ that node $i$ sends in repair of node $j$, and the repair data $S_{ji}$ that node $j$ sends in the repair of node $i$. 
Using this, we define for any set $A\subset \{1,\ldots, n\}$:
\begin{align}
S^{(j)}_{A} &\triangleq \{U_{i j}: i \in A\}\label{Definincoming}\\
U_{A}&\triangleq \{U_{i j}: (i, j)\in A, i\neq j\}\label{DefnPairwise}.
\end{align}

With these definitions in place, we can write the $(n-1)$ Type-II secrecy constraints for the first $(n-1)$ nodes as follows:
\begin{align}
I(F;  S_{[1:n-1]}\setminus \{S_{1}\})&=0\label{Securenot1}\\
I(F;  S_{[1:n-1]}\setminus \{S_{2}\})&=0\label{Securenot2}\\
&\hspace{5pt}\vdots\nonumber\\
I(F;  S_{[1:n-1]}\setminus \{S_{n-1}\})&=0\label{Securenotlast}.
\end{align}
where we have defined
\begin{align}
S_{[1:n-1]}\setminus \{S_{i}\}&\triangleq  (S_{1},\ldots, S_{i-1}, S_{i+1}, \ldots, S_{n-1}), 
\end{align}
for  $i=1,2,\ldots, (n-1)$. 

Using the constraint (\ref{Securenot1}) (i.e., secure repair of nodes $(2,3, \ldots, n-1)$), we have the following:
\begin{align}
H(F)&=H(F|S_{[1:n-1]}\setminus \{S_{1}\})\nonumber\\
&=H(F, S_{[1:n-1]}\setminus \{S_{1}\})- H(S_{[1:n-1]}\setminus \{S_{1}\})\label{T4PF1}.
\end{align}
Let us focus on the first term appearing in (\ref{T4PF1}):
\begin{align}
H(F, S_{[1:n-1]}\setminus \{S_{1}\})
&\leq H(F, W_{n}, S_{[1:n-1]}\setminus \{S_{1}\})\nonumber\\
&= H(W_{n}, S_{[1:n-1]}\setminus \{S_{1}\}) + H(F|W_{n}, S_{[1:n-1]}\setminus \{S_{1}\})\nonumber\\
&= H(W_{n}, S_{2}, S_{3}, \ldots, S_{n-1})+ H(F|W_{n}, S_{2}, S_{3}, \ldots, S_{n-1})\nonumber\\
&\leq H(W_{n}, S_{2}, S_{3}, \ldots, S_{n-1})+ H(F|W_{n}, W_{2}, W_{3}, \ldots, W_{n-1})\nonumber\\
&= H(W_{n}, S_{2}, S_{3}, \ldots, S_{n-1})\nonumber\\
&\leq H(W_{n}, S_{1}, S_{2}, S_{3}, \ldots, S_{n-1})\nonumber\\
&= H(W_{n}, U_{[1:n-1]}, S_{1}, S_{2}, S_{3}, \ldots, S_{n-1})\nonumber\\
&= H(W_{n}, U_{[1:n-1]}) + H(S_{1}, S_{2}, S_{3},\ldots, S_{n-1}|W_{n}, U_{[1:n-1]})\nonumber\\
&= H(W_{n}, U_{[1:n-1]})  + H(S_{n1}, S_{n2}, S_{n3},\ldots, S_{n(n-1)}|W_{n}, U_{[1:n-1]})\nonumber\\
&= H(W_{n}, U_{[1:n-1]}) \label{T4PF2},
\end{align}
where (\ref{T4PF2}) follows from the fact that $(S_{n1}, S_{n2},\ldots, S_{n(n-1)})$ are all functions of $W_{n}$. 

Next, we focus on the second term appearing in (\ref{T4PF1}):
\begin{align}
H(S_{[1:n-1]}\setminus \{S_{1}\})
&= H(S_{2}, \ldots, S_{n-1})\nonumber\\
&= H(S_{2},  \ldots, S_{n-1}, S_{21}, S_{2n}, \ldots, S_{(n-1)1}, S_{(n-1)n})\label{SPA}\\
&= H(U_{[1:n-1]}, S^{(n)}_{[2, 3, \ldots,n-1]})\label{SPB}\\
&= H(U_{[1:n-1]}) + H(S^{(n)}_{[2, 3, \ldots,n-1]}|U_{[1:n-1]})\label{T4PF3},
\end{align}
where (\ref{SPA}) follows from the fact that $W_{i}$  (and hence $(S_{i1}, S_{in})$) is a function of $S_{i}$.
Thus, as we have $S_{2}$, we can add $S_{21}, S_{2n}$; similarly, as we have $S_{i}$, we can add $(S_{i1}, S_{in})$
without increasing the entropy, for $i=2,3,\ldots,n$. Finally, (\ref{SPB}) follows by directly expanding all the terms $S_{2}, S_{3},\ldots, S_{n-1}$ and compactly expressing all the variables
by using the definitions of $U_{[1:n-1]}$ and $S^{(n)}_{[2, 3, \ldots, n-1]}$ which were defined in (\ref{DefnPairwise}) and
(\ref{Definincoming}).

Using (\ref{T4PF2}) and (\ref{T4PF3}) in (\ref{T4PF1}), we obtain
\begin{align}
H(F)&\leq H(W_{n}, U_{[1:n-1]}) - H(U_{[1:n-1]}) - H\left(S^{(n)}_{[2, 3, \ldots,n-1]}|U_{[1:n-1]}\right).
\end{align}
In summary, from the secure repair constraint of nodes  $\{1,\ldots, n-1\}\setminus \{1\}$, we have
\begin{align}
H(F)&\leq H(W_{n}, U_{[1:n-1]}) - H(U_{[1:n-1]})- H\left(S^{(n)}_{[1:n-1]\setminus \{1\}}|U_{[1:n-1]}\right).
\end{align}
Similarly, for the secure repair of nodes $\{1,\ldots, n-1\}\setminus \{i\}$, we can obtain
\begin{align}
H(F)&\leq H(W_{n}, U_{[1:n-1]}) - H(U_{[1:n-1]})  - H\left(S^{(n)}_{[1:n-1]\setminus \{i\}}|U_{[1:n-1]}\right).
\end{align}
There are total of $(n-1)$ such bounds for $i=1,2,\ldots, (n-1)$. Summing up these $(n-1)$ bounds, we obtain
\begin{align}
(n-1)H(F)
&\leq (n-1)H(W_{n}, U_{[1:n-1]}) - (n-1)H(U_{[1:n-1]})- \sum_{i=1}^{n-1} H\left(S^{(n)}_{[1:n-1]\setminus \{i\}}|U_{[1:n-1]}\right).\label{T4PF4}
\end{align}
We next focus on the summand appearing in (\ref{T4PF4}) for which we have the following inequality:
\begin{align}
\sum_{i=1}^{n-1} H\left(S^{(n)}_{[1:n-1]\setminus \{i\}}|U_{[1:n-1]}\right)
& \geq (n-2)H\left(S^{(n)}_{[1:n-1]}|U_{[1:n-1]}\right)\label{T4PF5},
\end{align}
which can be shown readily as follows:
\begin{align}
\sum_{i=1}^{n-1} H\left(S^{(n)}_{[1:n-1]\setminus \{i\}}|U_{[1:n-1]}\right)
&= \sum_{i=2}^{n-1} H\left(S^{(n)}_{[1:n-1]\setminus \{i\}}|U_{[1:n-1]}\right)
 + H\left(S^{(n)}_{[1:n-1]\setminus \{1\}}|U_{[1:n-1]}\right)\\
&= \sum_{i=2}^{n-1} H\left(S^{(n)}_{[1:n-1]\setminus \{i\}}|U_{[1:n-1]}\right)+ H\left(S^{(n)}_{[2, 3, \ldots, n-1] }|U_{[1:n-1]}\right)\\
&= \sum_{i=2}^{n-1} H\left(S^{(n)}_{[1:n-1]\setminus \{i\}}|U_{[1:n-1]}\right) + H\left(U_{2n}, U_{3n}, \ldots, U_{(n-1)n}|U_{[1:n-1]}\right)\\
&= \sum_{i=2}^{n-1} H\left(S^{(n)}_{[1:n-1]\setminus \{i\}}|U_{[1:n-1]}\right)+ H\left(U_{2n}|U_{[1:n-1]}\right) \nonumber \\
&\quad + H\left(U_{3n}|U_{2n}, U_{[1:n-1]}\right)+\ldots + H\left(U_{(n-1)n}|U_{2n}, U_{3n},\ldots, U_{(n-2)n},U_{[1:n-1]}\right)\nonumber \\
&\geq \sum_{i=2}^{n-1} H\left(S^{(n)}_{[1:n-1]\setminus \{i\}}|U_{[1:n-1]}\right)
+ H\left(U_{2n}|S^{(n)}_{[1:n-1]\setminus \{2\}}, U_{[1:n-1]}\right)\nonumber \\
&\quad+ H\left(U_{3n}|S^{(n)}_{[1:n-1]\setminus \{3\}} U_{[1:n-1]}\right)+\ldots + H\left(U_{(n-1)n}|S^{(n)}_{[1:n-1]\setminus \{(n-1)\}},U_{[1:n-1]}\right)\nonumber \\
&= (n-2)H\left(S^{(n)}_{[1:n-1]}|U_{[1:n-1]}\right).
\end{align}
This completes the proof for the bound (\ref{T4PF5}).

Using (\ref{T4PF5}) to further bound (\ref{T4PF4}), we obtain
\begin{align}
(n-1)H(F)
&\leq (n-1)H(W_{n}, U_{[1:n-1]}) - (n-1)H(U_{[1:n-1]})- (n-2) H\left(S^{(n)}_{[1:n-1]}|U_{[1:n-1]}\right)\nonumber\\
&= (n-1)H(W_{n}, U_{[1:n-1]}) - H(U_{[1:n-1]}) - (n-2) H\left(S^{(n)}_{[1:n-1]}, U_{[1:n-1]}\right)\nonumber\\
&= (n-1)H(W_{n}, U_{[1:n-1]}) - H(U_{[1:n-1]}) - (n-2) H\left(S^{(n)}_{[1:n-1]}, W_{n}, U_{[1:n-1]}\right)\label{T4PF6}\\
&\leq (n-1)H(W_{n}, U_{[1:n-1]}) - H(U_{[1:n-1]}) - (n-2) H\left(W_{n}, U_{[1:n-1]}\right)\nonumber\\
&= H(W_{n}, U_{[1:n-1]}) - H(U_{[1:n-1]})\nonumber\\
&\leq H(W_{n})+ H(U_{[1:n-1]}) - H(U_{[1:n-1]})\nonumber\\
&= H(W_{n})\nonumber\\
&\leq \alpha,\label{T4PF7}
\end{align}
where (\ref{T4PF6}) follows from the fact that $W_{n}$ is a function of $S^{(n)}_{[1:n-1]}$. To note this, we observe that 
$S^{(n)}_{[1:n-1]}$ among other variables,  consists of $(S_{1n}, S_{2n},\ldots, S_{(n-1)n})$, which is precisely the repair data for regenerating the information stored in node $n$ (i.e., $W_{n}$).

Hence, (\ref{T4PF7}) implies that $(n-1)H(F)\leq \alpha$, and hence we have the proof for the bound:
\begin{align}
\mathcal{B}^{S}_{II}\leq \frac{\alpha}{n-1}.
\end{align}

\end{document}